# Coordinated planning of European charging infrastructure and energy system for optimal V1G and V2G deployment

Francesco Sanvito[1], Francesco Lombardi[1], and Stefan Pfenninger-Lee[1]

[1]Faculty of Technology, Policy and Management (TPM), Delft University of Technology, Delft, the Netherlands
[*]Correspondence: f.sanvito@tudelft.nl

## ABSTRACT

Vehicle charging infrastructure targets in Europe currently rely on uniform benchmarks, overlooking the flexibility that could be offered by future smart charging (V1G) and Vehicle-to-Grid (V2G). To address this gap, we explicitly represent charging infrastructure and its costs in a cost-minimizing European energy system model, allowing uncontrolled, V1G, and V2G charging to compete. We find that V1G captures the majority of system cost savings (19–42 billion € year$^{-1}$; 2.2–4.5%) and substantially reduces infrastructure requirements, while V2G provides marginal savings up to 2.5 billion € year$^{-1}$ but generates substantial balancing market revenues (6.4 billion € year$^{-1}$). V2G deployment is most cost-effective in PV-dominated systems and limited-grid-expansion scenarios where combined solar and wind production is scarce. Charging requirements vary across countries, reflecting either utilization or flexibility maximization, indicating that uniform EU targets risk overestimating infrastructure needs in some regions while limiting the benefits of smart charging in others.

## INTRODUCTION

Tackling climate change requires a swift transition to low-emission energy systems. European policies, including REPowerEU(European Commission, 2022) and the Renewable Energy Directive (European Commission, 2021a), reinforce the continent's commitment to deep decarbonization by 2050(European Commission, 2019, 2021b). Within this broader transition, the transport sector remains a key challenge, accounting for over a quarter of total EU emissions (European Environment Agency, 2022). Passenger cars alone account for 43% of these emissions(Commission, 2024; European Commission, 2023).

The increasing electrification of transportation (Eurostat – European Commission, 2025) presents two key challenges. First, the deployment of low-carbon solutions should accelerate if climate targets are to remain within reach (Zeyen et al., 2025). Second, the growth in electric vehicles (EVs) demand should be coordinated to avoid unnecessary energy system costs (International Energy Agency, 2021; Mangipinto et al., 2022; Muratori, 2018; Helgeson and Peter, 2020; International Energy Agency, 2025) and to unlock additional system flexibility that might relieve pressure on the electricity grid (Crozier et al., 2020; Needell et al., 2023; Gunkel et al., 2020; Graabak et al., 2016).

Flexibility options mainly refer to (i) controlled charging (V1G), which shifts demand to align with system conditions and renewable generation, particularly midday solar (Wang et al., 2025; Powell et al., 2022), and (ii) bidirectional charging (V2G), which enables EVs to discharge electricity back to the grid but entails higher investment compared with unidirectional charging. Both V1G and V2G show potential to reduce reliance on other storage technologies and lower total energy system costs (Victoria et al., 2019; Zeyen et al., 2025; Kalweit et al., 2025; Bogdanov and Breyer, 2024; Syla et al., 2025).

Past studies on V1G and V2G report widely diverging energy system cost savings, ranging from none to 20% (Zeyen et al., 2025; Kalweit et al., 2025; Bogdanov and Breyer, 2024; Victoria et al., 2019; Syla et al., 2025; Brown et al., 2018b), and frequently rely on exogenous assumptions for V2G market value (ENTSO-E and ENTSOG, 2024; Syla et al., 2025, 2024) or conflate infrastructure costs with remuneration fees (Gunkel et al., 2020; Bogdanov and Breyer, 2024), limiting the assessment of potential economic benefits. Many models also neglect infrastructure investment as an independent decision (Bogdanov and Breyer, 2024; Zeyen et al., 2025; Kalweit et al., 2025; Guéret et al., 2024; Verzijlbergh et al., 2014), which prevents a proper evaluation of the competition between V1G and V2G. Additionally, restrictive assumptions on charging power and available battery capacity are often applied to avoid overestimating flexibility (Victoria et al., 2019; Syla et al., 2024; Kalweit et al., 2025; Gunkel et al., 2020; Zeyen et al., 2025; Brown et al., 2018a; Parajeles Herrera et al., 2026; Muessel et al., 2023; Miorelli et al., 2025; Wulff et al., 2020; Hanemann et al., 2017). Taken together, these limitations mean that no systematic assessment exists of whether V1G or V2G provides consistent system-level benefits, or under which energy system characteristics these flexibility options may be most effective.

The roll-out of V1G and V2G may affect charging requirements currently defined in the EU's Alternative Fuels Infrastructure Regulation (AFIR) (European Commission, 2025). Introducing investment costs tends to favor maximizing load factors and minimizing deployment, whereas exploiting lower electricity costs for charging can increase infrastructure needs. In addition, AFIR does not fully account for different energy system configurations or the long-term evolution of the EV fleet (Bernard et al., 2022a). It is therefore timely for policymakers to reconsider targets to avoid underused capacity, missed opportunities, and additional burdens on the energy transition.

Here, we evaluate the cost-optimal 2050 build-out of V1G and V2G charging infrastructure, the associated system-level benefits, the charging requirements for passenger EVs, and the potential V2G market revenues for each European country. Moreover, we identify the energy system conditions that may facilitate V2G deployment under different grid expansion scenarios. To achieve this, we extend the existing Sector-Coupled Euro-Calliope energy system optimization model (Pickering et al., 2022; Tröndle et al., 2019) with original mobility infrastructure modeling methods that we propose for the first time in the energy system planning literature. In particular, we endogenize charging and vehicle investment decisions, accounting for costs and country-specific EV battery constraints, which we obtain via a tailored version of the RAMP-mobility model (Mangipinto et al., 2022). Thanks to a temporal resolution of 1 hour, we capture the daily and seasonal flexibility patterns, while we represent each European country as a distinct node.

## V1G can significantly reduce peak charging requirements

We define scenarios along three dimensions: the level of charging flexibility (uncontrolled, V1G, V2G), cross-border transmission capacities (Limited, Moderate, and Unconstrained Grid Expansion, following the 2040 ENTSO-E projections (ENTSO-E, 2018)), and charging infrastructure costs (Cost-High, Cost-Base, Cost-Low, and Cost-Zero). Charging flexibility is represented by uncontrolled charging, V1G, and V2G, enabling the model to optimize its deployment. Each mode utilizes country-specific plug-in time series generated with the RAMP-mobility model (Sanvito, 2025), incorporating different plug-in probability functions based on the EV's battery state of charge. In addition to battery electric vehicles, the model also allows for alternative powertrains, including fuel cell electric vehicles (FCEVs) and internal combustion engine vehicles (ICEVs) running on diesel or synthetic diesel derivatives.

Figure 1 summarizes the Cost-Base results. Compared with uncontrolled charging, V1G substantially reduces peak charging requirements, from 281–284 GW to 73–94 GW, by shifting from uncoordinated plug-in behavior to price-responsive scheduling. In the uncontrolled case, limited EV charging flexibility leads the model to supply 5% of passenger mobility demand with FCEVs, since hydrogen production can be shifted in time.

Introducing V2G slightly increases the deployment of charging infrastructure to 76–116 GW, as additional charging is required to support later discharge during high-price hours. Only part of the charging network is upgraded with V2G capability, indicating that a full bidirectional deployment is not cost-effective. Additional transmission expansion reduces, but does not eliminate, the need for V2G infrastructure (Fig. 1i).

Across all non-zero cost scenarios, yearly V2G injections range from 1 to 89 TWh, requiring an additional charge of up to 14% to support bidirectional operation (Supplementary Figure 1). High V2G volumes place greater demands on vehicle connection time, as drivers must accommodate both extra charging and discharge, unless fast charging is widely available. In this regard, we investigate the changes in hourly and daily state-of-charge in Supplementary Note 1.

Overall, higher infrastructure costs or larger transmission expansion lead to lower V2G volumes, with infrastructure costs exerting the stronger influence.

## V1G drives most system cost reductions, with V2G adding limited savings but substantial market value

V1G reduces total energy system costs by 2.2–4.5% (19.4–42.4 billion € year$^{-1}$) (Supplementary Figure 2), excluding vehicle purchase costs and across all cost scenarios. Savings are highest when charging infrastructure costs are high, and transmission grid expansion is limited. V2G provides only marginal additional system benefits, ranging from zero to 2.5 billion € year$^{-1}$ (Fig. 2b, d, f).

From an EV owner's perspective, total charging costs comprise infrastructure expenditure, electricity purchases, and V2G revenues from market arbitrage. Expenditures and revenues are estimated using the model's dual variables as proxies for wholesale electricity prices (Brown et al., 2025), excluding country-specific taxes, charging point operators' profits, and labor costs. Moving from Uncontrolled to smart charging lowers electricity costs by 24-44%. Instead, moving from V1G to V2G increases electricity costs by 1-17%, but these additional costs are largely offset by V2G market revenues, which range from 0.2 to 6.4 billion € year$^{-1}$. As a result, the average charging price declines from 84-111 € MWh$^{-1}$ in the Uncontrolled case, in line with prior work (Lanz et al., 2022), to 45-59 € MWh$^{-1}$ in V1G and 41-59 € MWh$^{-1}$ in V2G.

At the country level, EV owners' potential savings from V2G can reach up to 66% of total charging costs compared to the V1G scenario, with higher performance in PV-rich countries (Fig. 2g). High V2G uptake increases aggregate injection volumes into the grid but reduces remuneration per energy unit, reflecting supply-driven price effects. Although EVs must remain connected for longer to secure equivalent revenues, the higher total V2G volume might enable broader participation, leading to dilution rather than crowding.

## Uniform charging infrastructure targets overlook country-specific requirements of V1G and V2G charging

The EU regulation on alternative fuels infrastructure (AFIR) sets a uniform target of 1.3 kW of public charging capacity per electric vehicle, irrespective of national energy system characteristics, the adoption of V1G or V2G charging, or EV penetration levels. Assessing the consistency of such uniform targets requires understanding how charging infrastructure requirements relate to their actual utilization across countries and charging scenarios.

We therefore compare charging infrastructure requirements, expressed as average charging power per vehicle, with the charging infrastructure load factor. These two quantities are mathematically linked and jointly characterize the deployment of charging infrastructure. In Figure 3, each country is represented by a curve tracing this relationship across utilization levels, with the curve shape determined by average EV energy use, distance traveled, and, under V2G, the volume of discharged energy relative to mobility demand. Further methodological details are provided in the Methods section.

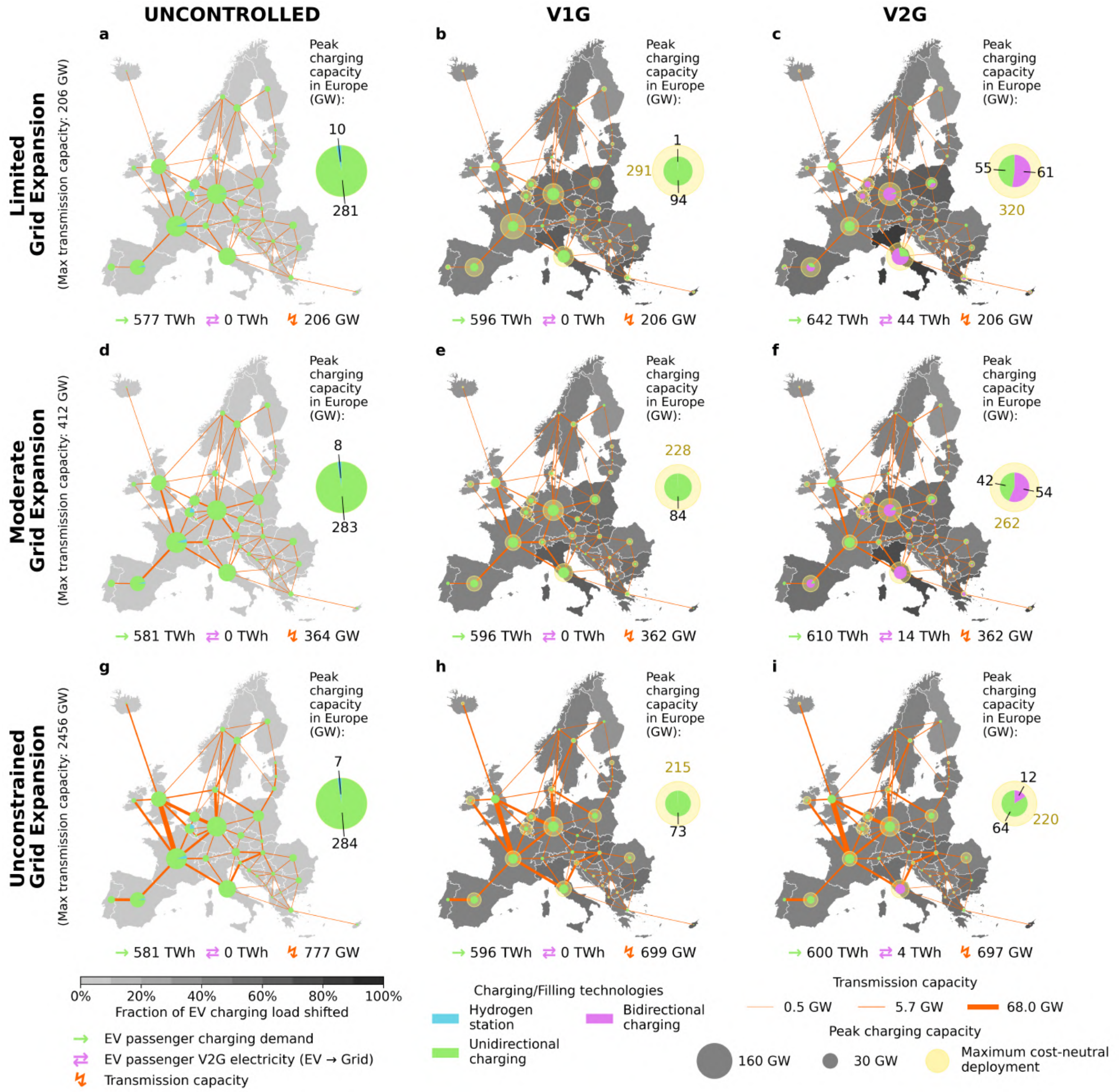

**Figure 1: Charging infrastructure deployment in the Cost-Base scenario.** The size of the circle in each country represents the installed charging capacity, while the color of the wedges indicates the type of charging or refueling infrastructure, distinguishing between hydrogen stations, unidirectional charging, and bidirectional charging. Yellow circles denote the maximum cost-neutral deployment, which is computed by dividing the difference in total energy system costs between the target scenario and the corresponding uncontrolled scenario, that is, the system benefits, by the charging infrastructure cost. This yields the additional infrastructure capacity that can be deployed without increasing total system costs. Transmission lines are also shown, with line width indicating transmission capacity. Additional metrics, including total charging demand, total V2G electricity injection relative to passenger EVs only, and total transmission capacity, are reported in each panel. Results refer exclusively to charging infrastructure deployed for passenger vehicles.

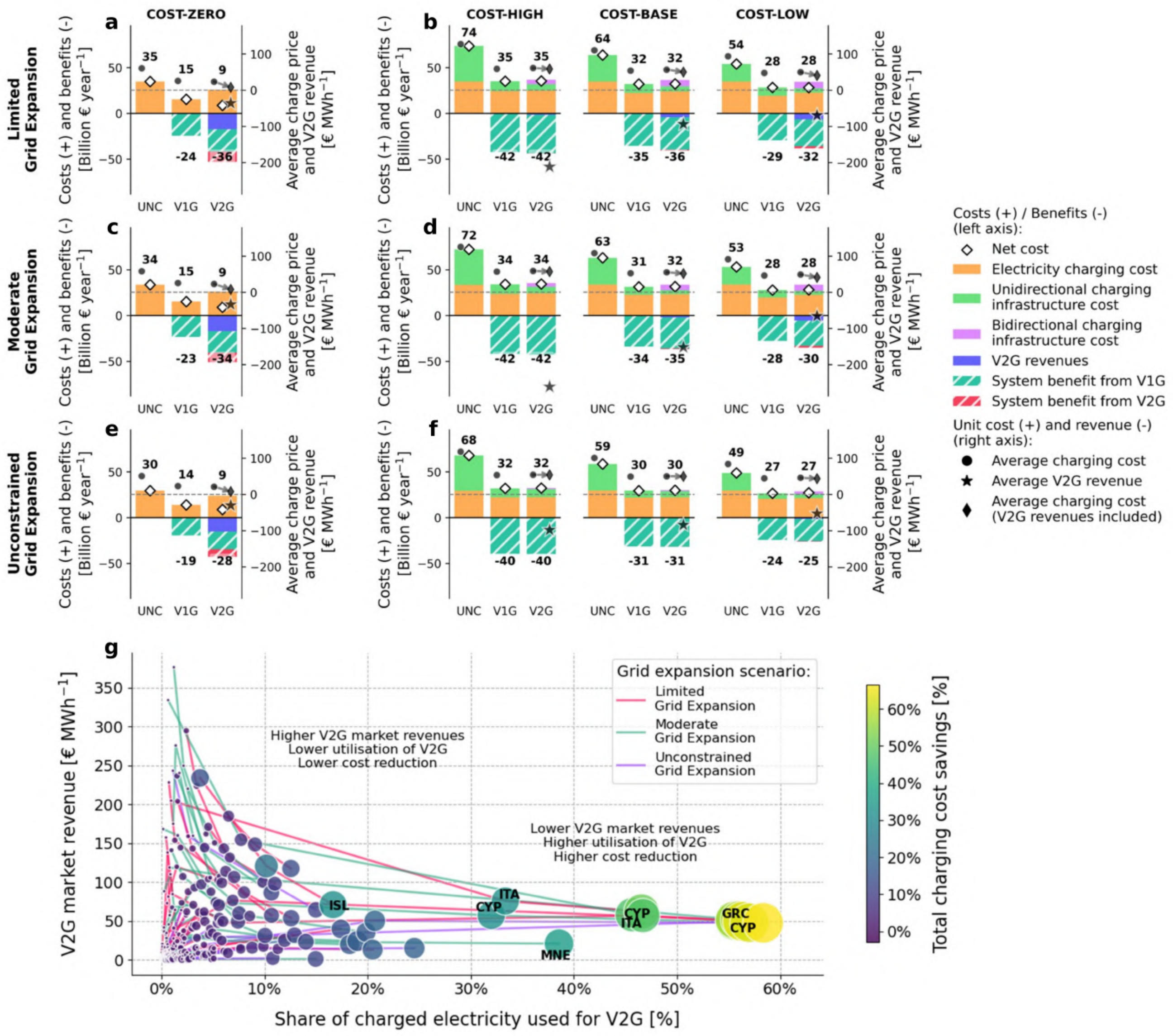

**Figure 2: Total charging costs and system-wide benefits.** Panels from (a) to (f) show the total cost-optimal charging costs for each scenario composed of three dimensions: grid expansion level, charging infrastructure flexibility level, and charging infrastructure cost estimate. Total charging costs (net cost) comprise infrastructure costs, total expenditure for electricity purchase, and V2G revenues. Total system-wide benefits (hatched fill) relative to V1G and V2G activation are reported as negative (as well as revenues). Total costs and total system-wide benefits figures are reported near each scenario bar. The total charging unit cost, with and without the effect of V2G market remuneration, is reported in each panel. Panel (g) shows the share of total charging cost savings in the V2G scenario compared to V1G for each modeled country, and across all non-zero charging infrastructure cost scenarios.

For unidirectional charging with identical mobility demand, this relationship is invariant to the temporal shape of charging, leading to identical curves for the Uncontrolled and V1G scenarios (light gray lines in Fig. 3a–b, d–e, and g–h). When V2G is enabled, discharged energy is included in the load-factor calculation, shifting the curves upward and weakening the inverse relationship between utilization and required capacity (Fig. 3c, f, and i). Overall, higher charging capacity is associated with lower average utilization but increased operational flexibility.

While these curves describe feasible combinations of utilization and capacity, results show that, in cost-optimal configurations, the infrastructure utilization factor is higher than in the Uncontrolled cases, thereby reducing total deployed charging capacity and keeping the energy system costs lower. Such outcomes can imply spatially concentrated charging deployment. To account for this tendency, we define a feasible range of charging infrastructure deployment using lower and upper bounds identified along each curve.

The lower bound corresponds to the cost-optimal energy system solution, in which V1G and V2G maximize system benefits with minimal deployment of charging infrastructure. The upper bound is defined by the maximum deployment level at which smart charging still delivers net system cost reductions relative to uncontrolled charging. Beyond this level, system benefits vanish; we therefore refer to this upper bound as the maximum cost-neutral deployment. Together, these bounds delineate the range of charging infrastructure deployment compatible with non-negative system benefits from smart charging.

Figure 3 illustrates the results for the Cost-Base scenario. When V1G and V2G are enabled, country-specific charging requirements span a wide range, from 0.17 to 2.81 kW vehicle$^{-1}$, with negligible differences between the two charging strategies. This wide dispersion indicates that charging infrastructure targets depend strongly on national energy system characteristics, challenging the applicability of a single EU-wide benchmark.

For example, Belgium benefits substantially from shifting from uncontrolled charging to V1G, as flexibility compensates for higher infrastructure costs. By contrast, countries such as the United Kingdom and Norway exhibit charging requirements below 0.5 kW vehicle$^{-1}$, reflecting energy systems dominated by wind and hydropower, where load shifting provides limited economic value. In absolute terms, total charging infrastructure provisioning for passenger EVs reaches 215–291 GW under V1G and 220–320 GW under V2G, depending on the grid expansion scenario (Fig. 1). Beyond these levels, additional deployment outweighs the benefits of smart charging and increases total energy system costs.

## Optimal V2G deployment is determined by PV prevalence in the energy mix

Optimal V2G deployment is primarily shaped by charging infrastructure costs, grid expansion, and the structure of the electricity mix. When charging infrastructure is expensive, and transmission capacity is available, V2G injections remain limited. In the Cost-High and Cost-Base scenarios, injections reach a maximum of 6 TWh in the Limited and Moderate Grid Expansion scenarios and approximately 24 TWh in the Unconstrained Grid Expansion scenarios. Lower charging infrastructure costs foster more V2G deployment, resulting in 89 TWh of grid injections. When charging infrastructure becomes inexpensive, and grid expansion remains limited, the generation mix becomes the primary determinant of V2G rollout. Under these conditions, solar prevalence becomes a prerequisite for significant deployment because large intra-day price spreads create profitable arbitrage opportunities for distributed storage. The competitiveness of V2G also depends on the marginal cost difference between V1G and V2G relative to alternative storage or dispatchable technologies, including options with lower cost or sunk investments such as pumped hydro storage or large hydro dams.

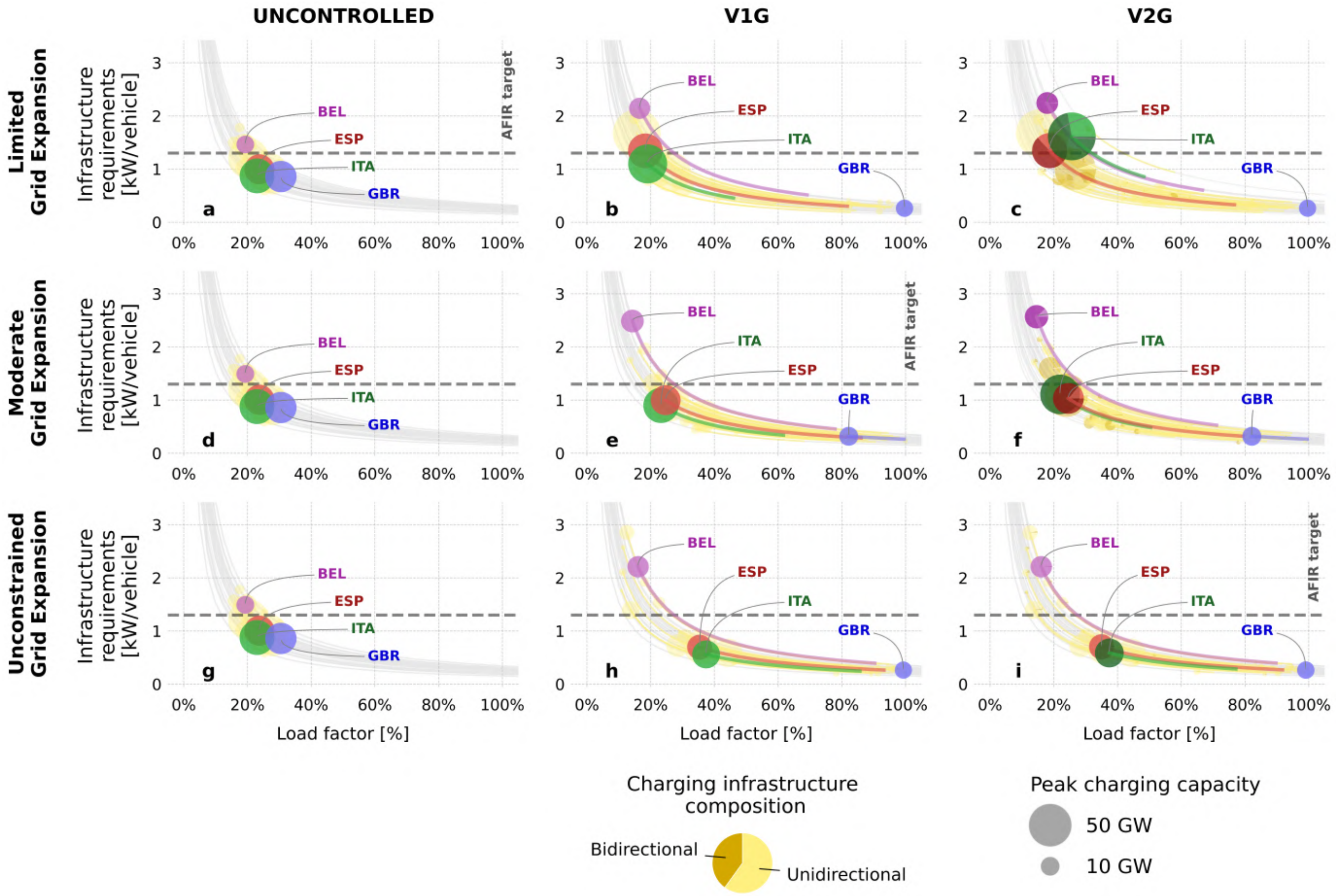

**Figure 3: Relationship between charging infrastructure requirements and the load factor.** Each circle represents a country, its size indicates the peak charging capacity level, and the colors indicate the type share of the charging infrastructure capacity. The yellow additional circle indicates the maximum theoretical level of charging infrastructure expansion, which minimizes infrastructure utilization and ensures at least positive energy system cost savings relative to the Uncontrolled scenario. Countries with the same level of specific EV charging consumption and traveled distance lie on the same gray curve, each with the same charging requirement coefficient (see Methods). The dashed line represents the infrastructure requirement target of 1.3 kW per electric vehicle set by the Alternative Fuel Infrastructure Regulation (AFIR) (European Commission, 2025).

As infrastructure costs continue to decline, V2G operations and the PV deployment become mutually reinforcing. In the Cost-Base scenario, PV deployment varies between −5% to +7%, whereas in the Cost-Low scenario, these effects become exclusively positive, ranging from 0 to +10%, accompanied by a decline in wind penetration (Supplementary Figure 16). In the model, transmission grid expansion, however, consistently enables greater PV deployment than V2G activation alone.

We assess the robustness of the V2G–PV relationship using a targeted Modeling to Generate Alternatives (MGA) analysis (Lombardi et al., 2025) applied to the Moderate Grid Expansion scenario in the Cost-Base and Cost-Low cases. When V2G deployment is maximized within a marginal 5% cost relaxation, net injections increase only marginally, and the generation mix remains essentially unchanged, indicating that the system has little need for additional V2G flexibility. In contrast, when PV capacity is minimized within the same cost relaxation, V2G injections decrease from 12 TWh to 7 TWh in the Cost-Base case and from 79 TWh to 1 TWh in the Cost-Low case. Lowering PV costs by 50% with a conventional sensitivity analysis further corroborates this dependency, increasing V2G activity from 12 TWh to 99 TWh in the Cost-Base scenario and from 79 TWh to 307 TWh in the Cost-Low scenario. This asymmetry shows that V2G responds to the opportunities created by solar generation, whereas the broader system exhibits limited sensitivity to changes in V2G availability. Sensitivity results are further discussed in Supplementary Note 2.

Smart charging also plays a central role in reducing system costs by displacing stationary battery storage. In the Cost-Base scenario, under Uncontrolled charging with limited transmission, storage requirements reach 874 GWh but fall to zero when both grid expansion and smart charging are enabled. Even without additional transmission capacity, smart charging yields greater reductions in storage requirements than grid expansion alone, as demonstrated by our analyses based on Hierarchical signed Shapley weighting (Supplementary Note 3), while V2G provides a smaller yet meaningful contribution. Transmission expansion primarily reduces reliance on dispatchable CHP and CCGT fueled by bio-derived or electric fuels. Moreover, V2G operations align renewable generation with flexible electric loads, increasing heat pump deployment by approximately 1% to 5%.

## V2G can provide flexibility when the combined solar and wind production is scarce and transmission expansion is limited

We now display the seasonal utilization pattern and the intraday operations of V2G for each modeled country (Fig. 5). The selected metrics are the maximum daily V2G production share for seasonal patterns and the V2G average utilization factor for intra-day operations.

Three seasonal usage patterns emerge. The reduced grid expansion level results in more consistent V2G utilization throughout the year. This happens in countries with PV-dominated energy mixes, such as Italy, Greece, and Cyprus. In the summer, EV charging and discharging replace the export of surplus PV generation.

The second pattern refers to countries where V2G usage increases only during the winter, when the lowest combined variable renewable generation rates occur. In these cases, V2G technologies act as peaking resources, discharging only a few hours per year, despite having a significant installed capacity. This occurs in both regions with balanced solar and wind resources and limited transmission, as well as in solar-dominated regions when grid expansion becomes less constrained. In the first case, the summer surplus is absorbed by flexible electrolyzers; in the second, spatial balancing

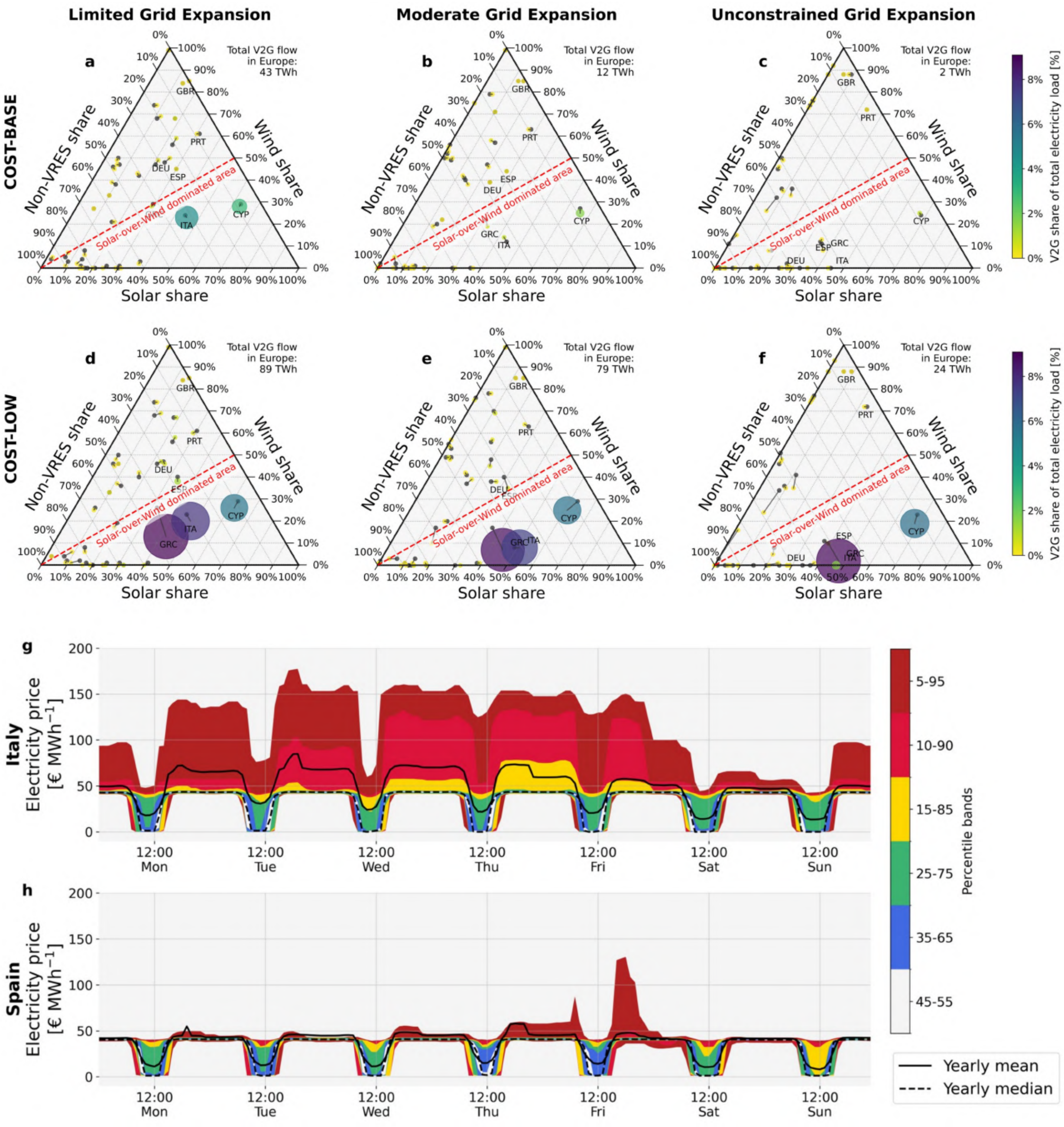

**Figure 4: Correlation between V2G adoption and energy system mix.** (a)-(f) Panels show ternary plots in which each circle represents a country. Grey circles represent V1G scenarios, colored circles represent V2G, and circles for the same countries are connected to each other. Size and color both refer to the share of V2G injection relative to each country's total gross electricity production. The red dashed line identifies the areas in which the solar share is greater than or equal to the wind share compared to the gross electricity production. (g)-(h) Panels show the electricity price spread over one year, reported to seven days, in the *Cost-Low* and *Moderate Grid Expansion* scenario for Italy and Spain. Italy achieves 50% PV penetration, whereas in Spain the wind share (35%) exceeds PV (33%). This difference results in an average maximum daily price spread of 47 € $MWh^{-1}$ in Italy, compared to 38 € $MWh^{-1}$ in Spain. This impacts the electricity injection from V2G, which in Italy reaches 56 TWh, compared to 2 TWh in Spain. Moreover, Spain has access to a greater capacity of cheap, dispatchable hydroelectric power plants that help mitigate electricity price fluctuations. Electricity price spreads and available flexibility options explain V2G adoption patterns.

via the grid reduces reliance on V2G operations (Supplementary Figure 17).

The third pattern occurs when other flexibility resources are more competitive compared to V2G. In these cases, bidirectional charging adoption is extremely limited or nonexistent, as pumped hydro or reservoir hydro, in combination with grid import and export when available, can ensure the country's energy system balance.

Overall, the costs of charging infrastructure modulate these patterns. High or baseline costs concentrate V2G dispatch in peak-price hours and maintain high load factors, whereas lower-cost infrastructure promotes more distributed and consistent use. In our study, V2G use is more concentrated in the summer months than in winter, when charging infrastructure costs are set to zero. This implies that charging infrastructure costs significantly impact bidirectional charging scheduling, especially on a seasonal basis.

Although seasonal V2G operation is intermittent, hourly operation follows a systematic pattern. EVs are requested to discharge electricity immediately before and after solar production peaks, which moderates the morning ramp and helps manage the evening peaks that remain after load shifting.

# DISCUSSION

Our study assesses the impact of optimally deploying V1G and V2G in a sector-coupled European energy system by 2050. Both V1G and V2G deployment levels are endogenously optimized by explicitly representing the charging infrastructure and its costs.

V2G infrastructure is consistently deployed across various scenarios, representing 11–56% of the total under varying cross-border transmission limits and infrastructure costs. Injection flows range from 0 to 89 TWh per year, a small share of the power sector balance, broadly consistent with TYNDP24 estimates of 55–80 TWh (ENTSO-E and ENTSOG, 2024). V2G utilization is triggered by large intraday electricity price spreads when cheaper dispatchable technologies are unavailable. Counterfactual analyses that exclude infrastructure costs show that omitting them yields unrealistically high V2G utilization and consequently overestimates the potential revenues of this technology (Supplementary Note 4).

In our study, V1G reduces system costs by 19.4–42.4 billion € per year (2.2–4.5%, excluding vehicle costs), while V2G contributes up to 2.5 billion € per year. V2G also generates gross revenues of up to 6.4 billion € per year. The magnitude of our cost-saving potential aligns with previous studies (Kern and Kigle, 2022; Kalweit et al., 2025), but our results highlight the major role of V1G, reflecting the importance of capturing charging infrastructure utilization. V1G captures most system-wide benefits, while V2G adds value only under specific system configurations. Under limited grid expansion and low infrastructure costs, V2G is favored in PV-dominated systems. PV deployment decreases when maximizing infrastructure load factors. With grid expansion and lower infrastructure costs, deployment dynamics reinforce each other, consistent with previous studies (Kern and Kigle, 2022; Powell et al., 2022; Guéret et al., 2024). The above limited and system-dependent value of V2G is corroborated by our targeted MGA and sensitivity analyses.

Previous research (Kalweit et al., 2025; Syla et al., 2025) reported either summer- or winter-dominated V2G patterns. Typically, only one pattern emerged per study, as infrastructure costs were often omitted or analyses focused on specific national systems. Our results show that both patterns occur, depending on PV availability, seasonal scarcity, or the availability of cheaper dispatchable resources, thereby reconciling earlier divergent findings. This aspect is crucial, as we show that V2G can help alleviate the winter gap problem and balance excess renewable production during summer months, reducing stress on grid connections.

This study does not assess V1G or V2G performance at the distribution-grid level and therefore omits explicit network constraints, which could reveal additional benefits. By including endoge-nous charging infrastructure costs, the model partially reflects utilization limits through load-factor optimization. V2G revenues are computed assuming full access to balancing markets, without accounting for potential losses or participation in other markets. The charging location is not considered, as the focus is on quantifying energy-system-consistent lower and upper bounds rather than local grid impacts. Limitations are further discussed in Supplementary Note 5.

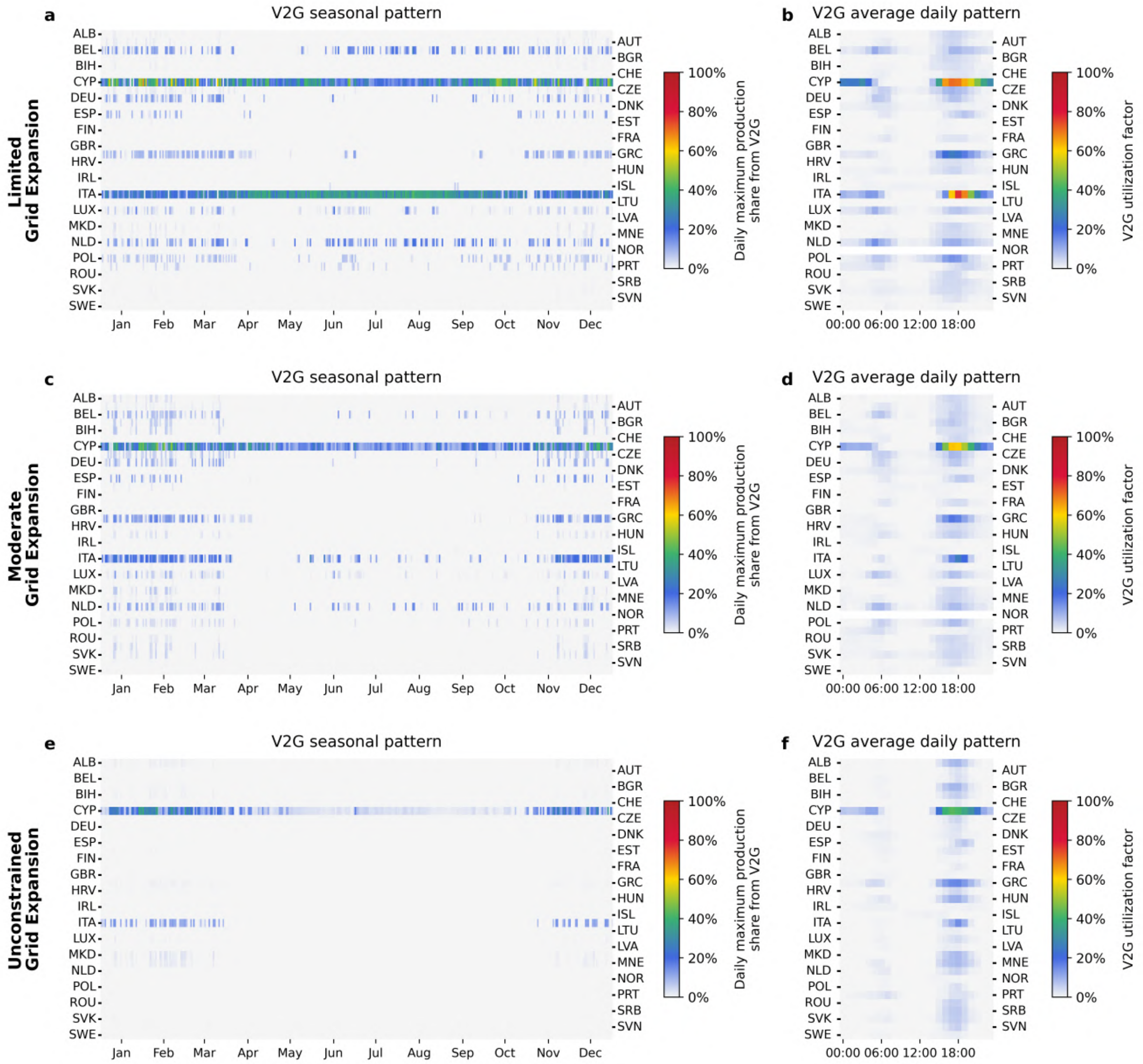

**Figure 5: V2G seasonal and daily utilization patterns.** (a), (c), (d), Panels show the daily maximum share of V2G contributing to the gross electric production. V2G is primarily used during winter periods when combined solar and wind production is lower, and the electricity price spread is driven by this scarcity. (b), (d), (f), Panels refer to the average V2G utilization factor over the year, showing a homogeneous time pattern across countries, which sees higher injections during the evening and, in a lighter fashion, during the morning before PV production ramps up. Results refer to the *Cost-Base* scenario.

Our spatially resolved model results raise questions about the uniformity of EU-wide charging infrastructure targets. The AFIR benchmark of 1.3 kW per vehicle ignores system value, national energy system differences, and EV penetration effects. While higher early-stage targets may support adoption, requirements decline as EV penetration and utilization increase (Bernard et al., 2022b). Fixed targets risk overdeployment in mature markets and under-provision where flexibility is most valuable. Planning should therefore be dynamic and aligned with system characteristics and expected V1G and V2G roles.

One important policy recommendation from our study is to prioritize V1G as a no-regret option since it unlocks the majority of system-level benefits. This can be achieved by allowing dynamic pricing and better aligning real-time coordination between vehicles and the grid. Moreover, V2G can be economically attractive in PV-dominated systems with significant intraday price volatility or limited alternative flexibility, including scenarios in which the grid expansion is delayed. These findings indicate that the cost-optimal deployment of V2G depends primarily on the characteristics of the underlying energy system. More broadly, the structural drivers identified here can inform national and sub-national assessments and show that coordinated V1G and V2G deployment not only reduces total system costs and relieves stress on other parts of the energy system, but also limits the need for additional energy infrastructure such as stationary batteries, dispatchable generation, or transmission expansion. By leveraging the flexibility already embedded in the EV fleet, such coordination can contribute to a more material-efficient energy transition.

# METHODS

## European energy system model

The model used in this study expands the model presented by Pickering et al. (Pickering et al., 2022) and Tröndle et al. (Tröndle et al., 2019), which represents the energy system across four main service sectors: *Heat*, which includes space heating, water heating, and cooking heat demands; *Fuels*, which covers bio-derived and synthetic fuels used in hard-to-electrify industrial processes, feedstock, aviation, and shipping; *Transport*, which encompasses motorcycles, passenger cars, buses, light-duty commercial vehicles, and heavy-duty freight vehicles; and *Electricity*, which accounts for building-level appliances and cooling, passenger and freight rail, and industrial processes.

We represent the following carriers in our model: electricity, hydrogen, $CO_2$, liquid and gaseous hydrocarbons (kerosene, methanol, diesel, and methane), solid fuels (residual biofuels, municipal waste, and biogas), low-temperature heat (including combined space heating, hot water, and cooking heat), and vehicle distance (disaggregated into five subcategories of road transport). These carriers can be produced, consumed, and converted by various technologies to meet end-use demands. In addition, low-temperature heat (at both household and district scales), hydrogen, electricity, and methane can be stored. Europe is assumed to be self-sufficient; therefore, imports to and exports from Europe are not considered.

Annual demand data are sourced from Eurostat (European Commission, 2020), JRC-IDEES (Mantzos et al., 2018), and Open Power System Data (Wiese et al., 2019). We follow the demand change assumptions outlined by Pickering et al. (Pickering et al., 2022). All demands are resolved in both time and space, except for liquid fuels such as diesel, kerosene, and methanol, which can be stored using existing infrastructure. Potential storage limitations for these fuels are outside the scope of this study.

Renewable generation capacity factors are defined at hourly resolution (Pfenninger and Staffell, 2016; Staffell and Pfenninger, 2016), and capacity limits are applied to account for deployment constraints (Tröndle et al., 2019; De Felice and Kavvadias, 2020). Technology costs and nameplate characteristics are primarily sourced from the Danish Energy Agency technology catalog (Danish Energy Agency, 2024), using projections for 2050.

We have extended the model with several additional features. Investments in power plants equipped with carbon capture systems can mitigate residual emissions from agriculture and industrial processes, in combination with geological storage, to support achieving a net-zero target by 2050. Consequently, the model allows the use of fossil fuels whose emissions can be captured through CCS and compensated by net-negative $CO_2$ technologies, such as direct air capture (DAC). There is a European carbon target of -200 Mton $CO_2$ to account for unavoidable emissions from industry and agriculture, and countries can compensate for other countries' emissions. Furthermore, the industrial demand for hydrogen and $CO_2$ is explicitly represented. Hydrogen and synthetic fuels can be distributed across regions at no additional cost, as we do not account for grid constraints or transportation costs in the exchange of fuels. While this assumption creates favorable conditions for powertrains competing with EVs, such as hydrogen or synthetic diesel, EVs still emerge as the preferred option, indicating that our results are robust to these modeling assumptions.

The analysis is implemented using the open-source energy system optimization framework Calliope (Pfenninger and Pickering, 2018) (version 0.6). The model operates at an hourly temporal resolution to capture the flexibility dynamics inherent to EV scheduling, while using a national (NUTS 0) spatial resolution. It minimizes total system costs, including vehicle purchase costs, within a perfect-foresight, single-year (2050) snapshot optimization framework.

Cross-border transmission capacity is based on ENTSO-E projections for 2040, defining the Limited Grid Expansion scenario (ENTSO-E, 2018). In the Moderate Grid Expansion scenario, constraints are relaxed by doubling the initial capacity of each transmission line. In the Unconstrained

Grid Expansion scenario, transmission capacity expansion is capped by a dynamic multiplier, but in practice, it is unconstrained.

We adopt the grid representation and related assumptions directly from Pickering et al. (Pickering et al., 2022). The initial high-voltage transmission network is based on the e-HIGHWAY 2050 project (e-Highway2050 Consortium, 2015), which provides simplified inter-regional power transfer capacities derived from detailed network analyses. It includes 48 planned or proposed upgrades from the 2018 ENTSO-E Ten-Year Network Development Plan (TYNDP) (ENTSO-E, 2018). These capacities act as lower bounds that can be further expanded. Since a purely linear representation of grid expansion may underestimate costs, differentiated expansion costs are applied based on actual expenditures from recent and planned projects, accounting for distance and terrain.

## Road transport sector

The model represents five main road transport segments: passenger cars, light-duty commercial vehicles, buses, heavy-duty vehicles, and motorcycles. Each segment includes three competing powertrains: diesel, battery-electric, and hydrogen. The only exception is motorcycles, for which only electric and liquid fuel options are considered. Internal combustion diesel engine powertrains can operate on fossil diesel or on hydrogen-derived and bio-based alternatives. Vehicle purchase costs are taken from various data sources (Danish Energy Agency, 2024; Burke et al., 2024). These costs are multiplied by the projected number of vehicles and then scaled to the peak demand to avoid underestimating the cost of replacing the entire vehicle fleet.

Electric powertrains are modeled through two distinct technologies: a storage technology and a conversion technology. The storage technology captures the ability to store electricity that can be discharged later, while the conversion technology converts stored electricity into transport service at a constant efficiency. For each electric technology, an additional carrier is introduced to represent the energy stored in vehicle batteries. To ensure consistent operation over time and to avoid unrealistic switching between technologies depending on renewable availability, we activate a Calliope constraint that enforces a constant powertrain share for each transport segment in all hours. The total electric storage capacity in each modeled region is limited by the number of vehicles multiplied by their assumed battery capacity. The number of vehicles is sourced from the JRC-IDEES database (Mantzos et al., 2017b). We do not assume vehicle numbers will grow, as they may vary with total driven distance under different conditions, such as model shifts, behavioral changes, or changes in household financial availability.

The annual service demands are expressed in vehicle-km and retrieved from multiple sources (Eurostat, 2025; Mantzos et al., 2017a). For each transport segment, service demand is represented with hourly resolution. Country-specific normalized time profiles for passenger cars, light-duty vehicles, and motorcycles are generated with the open-source stochastic mobility and charging profile generator RAMP-mobility (Mangipinto et al., 2022), which builds on RAMP software (Lombardi et al., 2024). Hourly activity profiles for buses and heavy-duty vehicles are derived from the BASt dataset (Bundesanstalt für Straßenwesen (BASt), 2025), normalized, and extended to all countries, as no better sources are available, while accounting for time zone differences as necessary.

## Scenario definition

The scenarios considered in this study explore three main dimensions: i) EV charging strategy, ii) grid expansion level, and iii) charging infrastructure costs. Each dimension is represented by multiple scenarios, described below. EV charging is represented using three scenarios. In the uncontrolled scenario, the charging profile is simulated using the RAMP-mobility model to mirror uncontrolled patterns. The share of electrified passenger cars is not fixed; instead, the model optimizes both the deployed charging infrastructure and the share of electrified vehicles in the fleet among candidate technologies, which include battery electric vehicles (EVs), fuel cell electric vehicles (FCEVs), and diesel vehicles running on conventional or synthetic fuels. In the V1G scenario, charging is scheduled according to electricity price signals to minimize system costs. Vehicle battery operation is constrained to safe state-of-charge (SoC) limits, and a time-dependent profile defines the maximum charge allowed, based on country-specific connection ratios derived from mobility patterns. Charging infrastructure costs are assumed to be identical to the uncontrolled case. In the V2G scenario, V1G and V2G infrastructure compete for deployment. Vehicle batteries are constrained for both charging and discharging, with connection rates modeled using a probability function that accounts for residual SoC, reflecting the willingness of vehicles to connect even when partially charged. Grid expansion is represented by three scenarios: limited expansion (maximum transmission capacity of 206 GW), moderate expansion (412 GW), and unconstrained expansion (2456 GW). Together, these charging and grid expansion scenarios enable a systematic analysis of the impacts of vehicle flexibility, fleet electrification, and transmission capacity on total system costs and electricity flows. More details about the selection of charging infrastructure costs are presented in the next section.

## Electric vehicles modeling

Vehicles within the same transport segment, namely passenger cars, light-duty vehicles, heavy-duty vehicles, buses, and motorcycles, are aggregated. Accordingly, battery capacity is aggregated by node and by transport segment. In this study, we improve the representation of electric vehicles by modeling them as the combination of two distinct technologies: a storage technology and a conversion technology. The storage technology aggregates the electric vehicle batteries in the fleet and stores electricity supplied by EV charging technologies. The conversion technology represents the vehicle powertrain, which converts electricity into traveled distance with an efficiency that depends on the vehicle type. A constraint links the deployed power capacity of the two technologies, ensuring that the diffusion of electric vehicles is proportional to each technology's available battery capacity. This modeling framework allows different powertrain technologies to contribute to transport demand while enforcing a constant share of each powertrain in every timestep. This avoids time-dependent distortions in the satisfaction of transport demand. We compute the maximum theoretical battery capacity for 100% electric vehicle penetration in each country and transport segment, which are then rescaled according to the endogenous EV penetration level (Supplementary Table 1). We assume battery capacities of 80 kWh for passenger vehicles, 150 kWh for light-duty vehicles, 15 kWh for motorcycles, 500 kWh for buses, and 1000 kWh for heavy-duty vehicles. Multiplying the vehicle battery capacity and the number of vehicles retrieved from Mantzos et al. (2018), we obtain the maximum levels of EV battery capacity. Cost assumptions are reported in Supplementary Note 8.

## Charging infrastructure modeling

Charging infrastructure is represented as a conversion technology that transforms electricity into a carrier that can be stored in vehicle batteries and subsequently used by the vehicle conversion technology to meet service demand. Two types of charging infrastructure are explicitly modeled: unidirectional and bidirectional. The unidirectional type converts electricity into another form that can be stored in electric vehicles, while the bidirectional type can operate in both directions. In the latter case, a round-trip efficiency is applied to account for conversion losses when electricity is re-injected into the grid.

In the Uncontrolled scenario, unidirectional charging follows the uncontrolled patterns from RAMP-mobility. The load factor is fixed to the corresponding normalized time series, while the infrastructure capacity remains subject to optimization.

In the V1G and V2G scenarios, charging infrastructure operation is left unconstrained, allowing the model to determine the optimal deployment of peak charging capacity. The flexibility of charging and discharging is limited by time series representing plug-in behavior, simulated using a modified version of RAMP-mobility (Sanvito, 2025). These normalized time series are used to cap the maximum charging and discharging rates of electric vehicle batteries, ensuring that energy flows remain consistent with plug-in patterns. The model does not constrain the state of charge of the electric vehicle batteries. The battery technology also includes upper and lower security bounds that ensure a minimum state of charge of 20% and a maximum of 80%.

Infrastructure costs vary depending on the application, type, number of installed chargers, and V2G capability (Lanz et al., 2022; Borlaug et al., 2020; Nicholas, 2019; Nelder and Rogers, 2019; Wood et al., 2023; Bernard et al., 2022a). To account for this variability, we select average costs for each vehicle type (Herbst et al., 2023; Bennett et al., 2022), distinguishing between V1G and V2G charging. Given the model's aggregation level and the uncertainty in future charging infrastructure costs, we conducted a sensitivity analysis to assess the impact of varying infrastructure costs and the relative cost difference between V1G and V2G on the system. We do not assume any exogenous remuneration fee for V2G participants. Moreover, we do not differentiate between charging power tiers of the charging infrastructure, and the cost is an average.

We also model hydrogen refueling stations for fuel-cell vehicles, assigning both investment and O&M costs (Bracci et al., 2024). The stations are assumed to dispense hydrogen in the gaseous phase.

Charging infrastructure and hydrogen station costs are further detailed in Supplementary Note 9.

In this study, we focus on passenger vehicles, as they have the longest parking times and the most variable mobility patterns compared to commercial vehicles. All other technologies are assumed to charge flexibly, representing a conservative scenario for evaluating the additional benefits of smart charging and bidirectional charging for private passenger vehicles in the energy system.

## V1G and V2G energy system benefits

We define system benefits as reductions in total energy system costs relative to the Uncontrolled scenario. The benefits of V1G are quantified as the difference in total system costs between the Uncontrolled and V1G scenarios. The incremental benefits of V2G are calculated by first determining the total cost reduction of the V2G scenario relative to the Uncontrolled case and subsequently subtracting the V1G benefits. This procedure isolates the portion of system cost savings attributable solely to V2G operation.

For country-specific benefit estimates, total system costs additionally include revenues and costs associated with the import and export of electricity and fuels, calculated by multiplying import or export quantities by the corresponding energy carrier shadow prices. Costs related to $CO_2$ capture and management and cross-border carbon compensation are excluded, as the carbon budget is set at the European level and these costs do not vary significantly across scenarios or countries.

## Charging requirements and utilization rate

Charging requirements are defined as the ratio of the charging infrastructure's peak capacity to the number of vehicles it serves. The two variables are linked by a coefficient termed the *charging requirements coefficient* (CRC), as shown in equation (1). The derivation of this coefficient is provided in equation (2).

$$\begin{matrix}\text{charging}\\\text{requirements}\end{matrix} = \frac{1}{\text{load factor}} \times \text{CRC} \tag{1}$$

$$
\begin{aligned}
\text{CRC} &= \frac{\text{charging}}{\text{requirements}} \times \text{load factor} = \\
&= \frac{\text{avg cons} \times \text{avg distance}}{\text{timesteps}} \times \left(1 + \frac{\text{discharge}_{\text{V2G}} \times \left(1 + \frac{1}{\eta_{V2G}}\right)}{\text{charge}_{\text{mob}}}\right)
\end{aligned}
\tag{2}
$$

The CRC is directly proportional to the average vehicle energy consumption ($kWh\ vehicle^{-1}$), the annual average traveled distance (km vehicle$^{-1}$ year$^{-1}$), the time horizon in hours, and a term accounting for electricity discharged through V2G operations. Specifically, the last term represents the ratio between the electricity discharged through V2G operations and the additional charging required to meet the mobility demand of the EV fleet, where $\eta_{\text{V2G}}$ denotes the round-trip efficiency of the V2G chargers. The sequence of calculations used to derive equation 2 is provided in Supplementary Note 6.

## Upper and lower charging requirement bounds

The model does not account for subnational spatial variation, leading to the optimization overestimating infrastructure utilization. Because charging station locations are not modeled, the optimization can associate charging activity with lower-capacity charging infrastructure, thereby maximizing the load factor. In reality, some stations would be underutilized due to their location, even though the aggregated charging and discharging profiles at the national level would remain broadly similar to the model-optimized profile.

To capture this effect, we progressively increase the maximum deployable charging capacity while keeping total charging energy constant. This approach enables us to identify the largest infrastructure rollout that still yields net system benefits under smart charging and V2G scenarios. The energy system, therefore, determines the point at which additional capacity no longer reduces total system costs. In this way, we can determine the upper bound of charging deployment or charging infrastructure requirements, while the lower bounds are represented by the results of the energy system optimization. Importantly, charging volumes and temporal patterns remain unchanged across scenarios; only the available peak capacity increases, which directly reduces the load factor.

As shown in Figure 3, these upper-bound charging requirement points lie on the same charging requirement curves as the reference scenarios, which correspond to the lower-bound levels. We therefore compute the resulting load factor using the same curve relationship and determine the corresponding maximum charging requirement by dividing country-specific system benefits by the average unit cost of charging infrastructure. When both V1G and V2G chargers are deployed, we use a weighted average of their respective infrastructure costs.

Since some countries exhibit negative system benefits, we first subtract the corresponding negative contributions from the total system benefits when calculating the country-specific benefits for V1G or V2G. The remaining positive system benefits are then redistributed among countries with positive system benefits only (Supplementary Note 7).

## Sensitivity analysis

We conduct several sensitivity analyses in addition to the scenarios that already test the robustness of our findings. We consider two main types of sensitivity. The first relates to the costs of charging infrastructure and photovoltaic technologies. In this case, we set the costs of V1G and V2G to the same value to test the relative penetration of V2G compared to V1G. This is relevant because, in the core model settings, V2G incurs a higher cost due to its enhanced bidirectional functionality. In the second sensitivity, we reduce the cost of photovoltaic technologies by 50%. This enables us to determine whether V2G adoption is associated with increased photovoltaic deployment across countries.

We also apply a targeted MGA analysis to check for deviations of our key findings near the cost-optimal regions (Lombardi et al., 2025). In detail, we implement V2G intensification and PV de-intensification while accounting for a 5% slack relative to the total energy system costs.

We find that equalizing V1G and V2G infrastructure costs substantially increases bidirectional charging deployment but has little effect on V2G discharge, suggesting that infrastructure cost differences primarily affect investment choices rather than operations. In contrast, reducing PV investment costs leads to lower charging infrastructure deployment and markedly higher V2G discharge, highlighting the strong dependence of V2G utilization on PV availability. Near-optimal analysis confirms this relationship, showing that increased V2G operation is effective only in systems with sufficiently high PV deployment. Results are reported and discussed in Supplementary Note 2.

# Data availability

The data and model instances used to generate the results of this study are available on Zenodo at: https://doi.org/10.5281/zenodo.18433372.

The results of this study are available on Zenodo at: https://doi.org/10.5281/zenodo.18431612.

# Code availability

The software version used in this study is available on GitHub at: https://github.com/FraSanvit/calliope/tree/0.6-v2g.

# References

Bennett, J., Mishra, P., Miller, E., Borlaug, B., Meintz, A., and Birky, A. (2022). Estimating the breakeven cost of delivered electricity to charge class 8 electric tractors. Technical Report NREL/TP-5400-82092, National Renewable Energy Laboratory. Accessed: 29 October 2025.

Bernard, M. R., Kok, I., Dallmann, T., and Ragon, P.-L. (2022a). Deploying charging infrastructure to support an accelerated transition to zero-emission vehicles. Technical report, International Council on Clean Transportation (ICCT). Accessed: 2 November 2025.

Bernard, M. R., Nicholas, M., Wappelhorst, S., and Hall, D. (2022b). A review of the afir proposal: How much power output is needed for public charging infrastructure in the european union? White paper, International Council on Clean Transportation (ICCT). March 17, 2022.

Bogdanov, D. and Breyer, C. (2024). Role of smart charging of electric vehicles and vehicle-to-grid in integrated renewables-based energy systems on country level. *Energy*, 301:131635.

Borlaug, B., Salisbury, S., Gerdes, M., and Muratori, M. (2020). Levelized cost of charging electric vehicles in the united states. *Joule*, 4(7):1470–1485.

Bracci, J., Koleva, M., and Chung, M. (2024). Levelized cost of dispensed hydrogen for heavy-duty vehicles. Technical Report NREL/TP-5400-88818, National Renewable Energy Laboratory. Accessed: 29 October 2025.

Brown, T., Hörsch, J., and Schlachtberger, D. (2018a). Pypsa: Python for power system analysis. *Journal of Open Research Software*, 6(1):4.

Brown, T., Neumann, F., and Riepin, I. (2025). Price formation without fuel costs: The interaction of demand elasticity with storage bidding. *Energy Economics*, 147:108483.

Brown, T., Schlachtberger, D., Kies, A., Schramm, S., and Greiner, M. (2018b). Synergies of sector coupling and transmission reinforcement in a cost-optimised, highly renewable european energy system. *Energy*, 160:720–739.

Bundesanstalt für Straßenwesen (BASt) (2025). Automatische dauerzählstellen – rohdaten (dz) for bundesfernstraßen. https://www.bast.de/DE/Publikationen/Daten/Verkehrstechnik/DZ.html?nn=427910. Accessed: 29 October 2025.

Burke, A. F., Zhao, J., and Fulton, L. M. (2024). Projections of the costs of light-duty battery-electric and fuel cell vehicles (2020–2040) and related economic issues. *Research in Transportation Economics*, 105:101440.

Commission, E. (2024). Eu transport in figures - statistical pocketbook 2024.

Crozier, C., Morstyn, T., and McCulloch, M. (2020). The opportunity for smart charging to mitigate the impact of electric vehicles on transmission and distribution systems. *Applied Energy*, 268:114973.

Danish Energy Agency (2024). Technology catalogue for commercial freight and passenger transport. Accessed: 2025-09-24.

De Felice, M. and Kavvadias, K. (2020). Energymodelling-toolkit/hydro-power-database: Jrc hydro-power database, release 07. https://zenodo.org/record/5215920. Zenodo.

e-Highway2050 Consortium (2015). Modular development plan of the pan-european transmission system 2050. https://docs.entsoe.eu/baltic-conf/bites/www.e-highway2050.eu/e-highway2050/. Accessed: 2025-10-28.

ENTSO-E (2018). Tyndp 2018 scenario report. Technical report, European Network of Transmission System Operators for Electricity (ENTSO-E). Accessed: 2025-04-27.

ENTSO-E and ENTSOG (2024). Tyndp 2024 scenarios report. European Network of Transmission System Operators for Electricity (ENTSO-E) and for Gas (ENTSOG). Downloaded from 2024 ENTSO-E/ENTSOG Scenarios page.

European Commission (2019). The european green deal. https://eur-lex.europa.eu/legal-content/EN/TXT/?uri=CELEX:52019DC0640. Communication from the Commission, COM(2019) 640 final.

European Commission (2020). Eurostat database. https://ec.europa.eu/eurostat/data/database. Accessed: 2025-10-28.

European Commission (2021a). Proposal for a directive of the european parliament and of the council on the promotion of the use of energy from renewable sources (recast). https://eur-lex.europa.eu/legal-content/EN/TXT/?uri=CELEX%3A52021PC0575. Proposal, COM(2021) 575 final.

European Commission (2021b). 'fit for 55': Delivering the eu's 2030 climate target on the way to climate neutrality. https://eur-lex.europa.eu/legal-content/EN/TXT/?uri=CELEX%3A52021DC0550. Communication from the Commission, COM(2021) 550 final.

European Commission (2022). Repowereu: Joint european action for more affordable, secure and sustainable energy. https://eur-lex.europa.eu/legal-content/EN/TXT/?uri=CELEX%3A52022DC0221. Communication from the Commission, COM(2022) 221 final.

European Commission (2023). Co2 emission performance standards for cars and vans. https://climate.ec.europa.eu/eu-action/transport/road-transport-reducing-co2-emissions-vehicles/co2-emission-performance-standards-cars-and-vans_en. Accessed: 2025-04-26.

European Commission (2025). Alternative fuels infrastructure regulation (afir). https://transport.ec.europa.eu/transport-themes/clean-transport/alternative-fuels-sustainable-mobility-europe/alternative-fuels-infrastructure_en. Accessed: 2025-11-24.

European Environment Agency (2022). Sustainability of europe's mobility systems – climate impacts. https://www.eea.europa.eu/en/analysis/publications/sustainability-of-europes-mobility-systems/climate. Accessed: 2025-04-26.

Eurostat (2025). Road transport vehicles by category — table road_tf_veh. https://ec.europa.eu/eurostat/databrowser/view/road_tf_veh/default/table?lang=en&category=road.road_tf. Accessed: 2 November 2025.

Eurostat – European Commission (2025). Stock of electric vehicles by category and nuts 2 region (dataset: tran_r_elvehst). Data as of May2025; accessed: 2025-07-26.

Graabak, I., Wu, Q., Warland, L., and Liu, Z. (2016). Optimal planning of the nordic transmission system with 100 *Energy*, 107:648–660.

Gunkel, P. A., Bergaentzlé, C., Græsted Jensen, I., and Scheller, F. (2020). From passive to active: Flexibility from electric vehicles in the context of transmission system development. *Applied Energy*, 277:115526.

Guéret, A., Schill, W.-P., and Gaete-Morales, C. (2024). Impacts of electric carsharing on a power sector with variable renewables. *Cell Reports Sustainability*, 1(11). Publisher: Elsevier.

Hanemann, P., Behnert, M., and Bruckner, T. (2017). Effects of electric vehicle charging strategies on the german power system. *Applied Energy*, 203:608–622.

Helgeson, B. and Peter, J. (2020). The role of electricity in decarbonizing european road transport – development and assessment of an integrated multi-sectoral model. *Applied Energy*, 262:114365.

Herbst, I., Barth, J., Daneshgar, G., Furrer, R., Heer, L., Hickethier, M., Mathis, U., Rosser, S., Trottmann, M., Vischer, M., Wahl, H., and Walter, S. (2023). Laden im quartier: Informationssammlung zur elektromobilität für gemeinden. https://pubdb.bfe.admin.ch/de/publication/download/12068. Accessed: 29 October 2025; Prepared for *Roadmap Elektromobilität 2025*, Bundesamt für Energie (BFE), Switzerland.

International Energy Agency (2021). Global ev outlook 2021. https://www.iea.org/reports/global-ev-outlook-2021. IEA, Paris. Licence: CC BY 4.0.

International Energy Agency (2025). Global ev outlook 2025. https://www.iea.org/reports/global-ev-outlook-2025. IEA, Paris. Licence: CC BY 4.0.

Kalweit, S., Zeyen, E., and Victoria, M. (2025). Endogenous transformation of land transport in europe for different climate targets.

Kern, T. and Kigle, S. (2022). Modeling and evaluating bidirectionally chargeable electric vehicles in the future european energy system. *Energy Reports*, 8(2):694–708.

Lanz, L., Noll, B., Schmidt, T. S., and Steffen, B. (2022). Comparing the levelized cost of electric vehicle charging options in Europe. *Nature Communications*, 13(1):5277.

Lombardi, F., Duc, P.-F., Tahavori, M. A., Sanchez-Solis, C., Eckhoff, S., Hart, M. C., Sanvito, F., Ireland, G., Balderrama, S., Kraft, J., Dhungel, G., and Quoilin, S. (2024). Ramp: stochastic simulation of user-driven energy demand time series. *Journal of Open Source Software*, 9(98):6418.

Lombardi, F., van Greevenbroek, K., Grochowicz, A., Lau, M., Neumann, F., Patankar, N., and Vågerö, O. (2025). Near-optimal energy planning strategies with modeling to generate alternatives to flexibly explore practically desirable options. *Joule*, 9(11). Publisher: Elsevier.

Mangipinto, A., Lombardi, F., Sanvito, F. D., Pavičević, M., Quoilin, S., and Colombo, E. (2022). Impact of mass-scale deployment of electric vehicles and benefits of smart charging across all european countries. *Applied Energy*, 312:118676.

Mantzos, L., Rozsai, M., Matei, N. A., Mulholland, E., Tamba, M., and Wiesenthal, T. (2018). Jrc-idees 2015.

Mantzos, L., Wiesenthal, T., Matei, N., Tchung-Ming, S., Rozsai, M., Russ, P., and Soria Ramirez, A. (2017a). Jrc-idees: Integrated database of the european energy sector: Methodological note. Technical report, Joint Research Centre (Seville site).

Mantzos, L., Wiesenthal, T., Matei, N., Tchung-Ming, S., Rózsai, M., Russ, H., and Soria Ramirez, A. (2017b). Jrc-idees: Integrated database of the european energy sector: Methodological note. Technical Report JRC108244, Publications Office of the European Union, Luxembourg.

Miorelli, F., Wulff, N., Fuchs, B., Gils, H. C., and Jochem, P. (2025). venco.py: A python model to represent the charging flexibility and vehicle-to-grid potential of electric vehicles in energy system models. *Journal of Open Source Software*, 10(108):6896.

Muessel, J., Ruhnau, O., and Madlener, R. (2023). Accurate and scalable representation of electric vehicles in energy system models: A virtual storage-based aggregation approach. *iScience*, 26(10). Publisher: Elsevier.

Muratori, M. (2018). Impact of uncoordinated plug-in electric vehicle charging on residential power demand. *Nature Energy*, 3(3):193–201.

Needell, Z., Wei, W., and Trancik, J. E. (2023). Strategies for beneficial electric vehicle charging to reduce peak electricity demand and store solar energy. *Cell Reports Physical Science*, 4(3):101287.

Nelder, C. and Rogers, E. (2019). Reducing ev charging infrastructure costs. Technical report, Rocky Mountain Institute (RMI). Accessed: 2 November 2025.

Nicholas, M. (2019). Estimating electric vehicle charging infrastructure costs across major u.s. metropolitan areas. Technical report, International Council on Clean Transportation (ICCT). Accessed: 2 November 2025.

Parajeles Herrera, M., Garrison, J., and Hug, G. (2026). The role of unidirectional charging flexibility in the planning and operation of the future swiss electricity system. *Applied Energy*, 403:127078.

Pfenninger, S. and Pickering, B. (2018). Calliope: a multi-scale energy systems modelling framework. *Journal of Open Source Software*, 3(29):825.

Pfenninger, S. and Staffell, I. (2016). Long-term patterns of european pv output using 30 years of validated hourly reanalysis and satellite data. *Energy*, 114:1251–1265.

Pickering, B., Lombardi, F., and Pfenninger, S. (2022). Diversity of options to eliminate fossil fuels and reach carbon neutrality across the entire european energy system. *Joule*, 6(6):1253–1276.

Powell, S., Cezar, G. V., Min, L., Azevedo, I. M. L., and Rajagopal, R. (2022). Charging infrastructure access and operation to reduce the grid impacts of deep electric vehicle adoption. *Nature Energy*, 7(10):932–945.

Sanvito, F. (2025). Ramp-mobility (version v2g_output). https://github.com/FraSanvit/RAMP-mobility/tree/V2G_output. Accessed: 2025-04-27.

Staffell, I. and Pfenninger, S. (2016). Using bias-corrected reanalysis to simulate current and future wind power output. *Energy*, 114:1224–1239.

Syla, A., Parra, D., and Patel, M. K. (2025). Assessing flexibility from electric vehicles using an open-source energy system model: trade-offs between smart charging, vehicle-to-grid and an extensive charging infrastructure. *Energy*, 326:136236.

Syla, A., Rinaldi, A., Parra, D., and Patel, M. K. (2024). Optimal capacity planning for the electrification of personal transport: The interplay between flexible charging and energy system infrastructure. *Renewable and Sustainable Energy Reviews*, 192:114214.

Tröndle, T., Pfenninger, S., and Lilliestam, J. (2019). Home-made or imported: on the possibility for renewable electricity autarky on all scales in europe. *Energy Strategy Reviews*, 26:100388.

Verzijlbergh, R., Brancucci Martínez-Anido, C., Lukszo, Z., and de Vries, L. (2014). Does controlled electric vehicle charging substitute cross-border transmission capacity? *Applied Energy*, 120:169–180.

Victoria, M., Zhu, K., Brown, T., Andresen, G. B., and Greiner, M. (2019). The role of storage technologies throughout the decarbonisation of the sector-coupled european energy system. *Energy Conversion and Management*, 201:111977.

Wang, Z., Sasse, J.-P., and Trutnevyte, E. (2025). Home or workplace charging? spatio-temporal flexibility of electric vehicles within swiss electricity system. *Energy*, 320:135452.

Wiese, F., Schlecht, I., Bunke, W.-D., Gerbaulet, C., Hirth, L., Jahn, M., Kunz, F., Lorenz, C., Mühlenpfordt, J., Reimann, J., and Schill, W.-P. (2019). Open power system data—frictionless data for electricity system modelling. *Applied Energy*, 236:401–409.

Wood, E., Borlaug, B., Moniot, M., Lee, D.-Y., Ge, Y., Yang, F., and Liu, Z. (2023). The 2030 national charging network: Estimating u.s. light-duty demand for electric vehicle charging infrastructure. Technical Report NREL/TP-5400-85654, National Renewable Energy Laboratory (NREL). Accessed: 2 November 2025.

Wulff, N., Steck, F., Gils, H. C., Hoyer-Klick, C., van den Adel, B., and Anderson, J. E. (2020). Comparing power-system and user-oriented battery electric vehicle charging representation and its implications on energy system modeling. *Energies*, 13(5).

Zeyen, E., Kalweit, S., Victoria, M., and Brown, T. (2025). Shifting burdens: how delayed decarbonisation of road transport affects other sectoral emission reductions. *Environmental Research Letters*, 20(4):044044.

# Acknowledgements

We acknowledge the use of computational resources of the DelftBlue supercomputer, provided by Delft High Performance Computing Centre (https://www.tudelft.nl/dhpc).

This work was performed within the PATHFNDR project, which is sponsored by the Swiss Federal Office of Energy's SWEET programme under Grant Number SI/502259.

F.S. acknowledges Bryn Pickering for developing the post-processing script for Calliope results and the initial version of the map-plotting script, as well as for insightful discussions on transport and energy-system modelling issues.

# SUPPLEMENTARY INFORMATION

# Supplementary Note 1: Electric vehicle fleet State-of-Charge

We conduct a reality check to assess battery utilization across scenarios in the electric vehicle (EV) fleet, with particular attention to cases in which vehicle-to-grid (V2G) operation increases battery cycling through grid injections and the associated additional charging demand. The analysis focuses on two aspects: (i) hourly battery utilization and (ii) daily changes in the state of charge (SoC).

Supplementary Figure 1 shows that, even in the most unfavorable case from a battery utilization perspective (Cost-Low, Limited Grid Expansion, and V2G), the hourly change in SoC remains below 2% of the total battery capacity of the EV fleet. Holding all other assumptions constant, the Cost-Zero case exhibits substantially higher utilization, with peak hourly SoC change reaching approximately 5%, more than doubling hourly battery use relative to the Cost-Low case.

Supplementary Figure 2 shows that daily SoC changes can reach up to ±10%. In several countries, systematic differences between weekdays and weekends are observed. During weekdays, the balance between EV charging and EV electricity consumption is predominantly negative, whereas on weekends it becomes positive, replenishing the EV fleet's batteries. This pattern reflects a combined effect of lower passenger vehicle activity on weekends compared to weekdays and slightly higher charging levels, which may be driven by lower electricity prices.

# Supplementary Note 2: Sensitivity analysis

Alongside the set of scenarios developed, we conduct an additional sensitivity analysis structured as follows.

First, we test the robustness of the results related to V2G infrastructure deployment by assigning the same infrastructure cost to V1G and V2G, thereby eliminating the cost difference associated with bidirectional functionality. The results are reported in Table 1. This modification is applied to two scenarios with moderate grid expansion, namely the Cost-Base and Cost-Low cases. We find that although bidirectional charging deployment increases substantially and almost entirely replaces unidirectional smart infrastructure, V2G discharge volumes remain largely unchanged. In this case, the additional cost associated with bidirectional capability strongly influences the type of infrastructure deployed, but not its operation. In contrast, reducing PV investment by 50% leads to the opposite outcome: lower infrastructure deployment and a pronounced increase in V2G operations. V2G discharge rises from 12 TWh to 99 TWh in the Cost-Base case (Supplementary Figure 3c) and from 79 TWh to 307 TWh in the Cost-Low case (Supplementary Figure 3d). Lower PV costs favor PV deployment and, consequently, increase V2G utilization.

Second, we perform a Modeling-to-Generate-Alternatives analysis (Lombardi et al., 2025) to assess deviations from the cost-optimal solution in the near-optimal region. Specifically, we de-intensify PV deployment and intensify V2G operation by allowing a 5% slack in total energy system costs, including vehicle costs.

When PV installation is minimized, V2G discharge decreases substantially, by 42% in the Cost-Base scenario and by 82% in the Cost-Low scenario (Supplementary Figures 3e and 3f). Conversely, when V2G operation is intensified, V2G discharge increases only in countries that already exhibit V2G penetration (Supplementary Figures 3g and 3h). Overall, these results indicate that higher PV deployment drives increased V2G operation, while intensified V2G operation is effective only when PV capacity is sufficiently deployed.

# Supplementary Note 3: Quantifying the relative impact of each scenario dimension

To quantify the relative influence of each scenario dimension on model outcomes, we computed hierarchical signed Shapley-type weights. For each technology $g$, we first aggregated the absolute changes in the output variable (e.g., installed capacity) across scenario dimensions $d$ to determine their relative magnitudes. Within each dimension, we further disaggregated the effects by subtype $u$, corresponding to specific scenario transitions (for example, within the charging dimension: uncontrolled→V1G and V1G→V2G), and by the sign of their contribution (positive or negative). Each subtype effect can be decomposed into its constituent modifications $\Delta_{g,t}$, such that

$$a_{g,d,u} = \sum_{t \in T_{g,d,u}} \Delta_{g,t}, \qquad A_{g,d} = \sum_{u \in \mathcal{U}_{g,d}} a_{g,d,u} = \sum_{u \in \mathcal{U}_{g,d}} \sum_{t \in T_{g,d,u}} \Delta_{g,t}. \tag{1}$$

These subtype-level effects were normalized within each dimension to capture their internal composition. The resulting hierarchical absolute weights were then combined with the sign of each subtype's net effect and normalized such that the sum of absolute weights equals one for each technology. This yields signed weights that express the proportional contribution of each dimension and subtype to the overall modeled difference.

We define the relative share of each targeted dimension $d^*$ and the normalized contribution of each targeted subtype $u^*$ within that dimension according to equation (2).

$$p_{g,d^*} = \frac{A_{g,d^*}}{\sum_d A_{g,d}}, \qquad q_{g,d^*,u^*} = \frac{a_{g,d^*,u^*}}{\sum_{u \in \mathcal{U}_{g,d^*}} a_{g,d^*,u}}, \tag{2}$$

where $A_{g,d}$ and $a_{g,d,u}$ are defined as above, representing the total absolute change for each dimension and subtype, respectively.

The final signed, normalized hierarchical weight is then given by equation (3).

$$w^{\mathrm{norm}}_{g,d^*,u^*} = \mathrm{sgn}\Big(a_{g,d^*,u^*}\Big)\, p_{g,d^*}\, q_{g,d^*,u^*}, \tag{3}$$

where $\mathrm{sgn}(x) = +1$ if $x \geq 0$ and $-1$ otherwise. Here, $d^*$ and $u^*$ indicate the targeted dimension and subtype for which the weight is computed, while $d$ and $u$ in the denominators denote summation over all dimensions and all subtypes within each dimension, respectively. This normalization ensures that $\sum_{d^*,u^*} |w^{\mathrm{norm}}_{g,d^*,u^*}| = 1$ for each technology $g$, yielding directly comparable relative contributions of each dimension and subtype across technologies.

The methodology can be applied equivalently to changes in installed capacity or energy flows. In the following, we present an analysis comparing different charging strategies and grid expansion cases under the Cost-Low scenario.

This methodology is particularly suited to investigating the impact of V1G and V2G on the deployment of dispatchable technologies, rather than on the role of transmission expansion, as both charging strategies and grid expansion contribute to increasing system flexibility. Overall, dispatchable capacity decreases across all Cost-Low scenarios by 12% to 50%. According to the implemented Shapley methodology, V1G accounts for the largest share of this reduction, followed by transmission expansion and V2G.

Disaggregating dispatchable technologies reveals that battery storage is the most widely deployed option and is primarily affected by V1G activation, with grid expansion and V2G having comparable effects. In contrast, reductions in combined-cycle gas turbines and combined heat and power technologies are mainly driven by the expansion of the transmission grid from the Moderate to the Unconstrained case. This highlights the potential of V1G to substantially reduce the deployment of dispatchable generation in scenarios where transmission capacity remains constrained (Supplementary Figure 4).

Supplementary Figures 4g–i illustrate cross-sectoral interactions between the power and heat sectors. Reduced flexibility in the power sector leads to lower combined heat and power capacity, thereby creating room for increased heat pump deployment. While the decline in combined heat and power capacity is primarily driven by grid expansion, heat pump penetration increases due to both V1G and V2G adoption. In Supplementary Figure 5, we show the results for the energy flows.

The magnitude of these effects depends on assumptions about charging infrastructure costs. For this reason, we also report changes in capacity deployment that explicitly account for cost variations.

# Supplementary Note 4: Impacts of neglecting charging infrastructure costs

Assigning costs to charging infrastructure not only affects total system costs but also influences the scheduling of electric vehicles (EVs). While adding charging infrastructure costs to vehicle costs does not directly alter the shape of EV charging profiles, it substantially affects charging event scheduling, infrastructure deployment, and utilization.

First of all, the charging profiles in the Cost-Zero scenarios exhibit strong temporal alignment with PV production, leading to pronounced midday charging load peaks, as shown in Supplementary Figure 6 (Cost-Zero) and Supplementary Figure 7 (Cost-Base), both in the Moderate Grid Expansion scenario. Such results completely disregard the potential impacts of pronounced midday charging peaks that may overload distribution grids.

In other words, EV charging decisions are driven solely by electricity prices, without accounting for the required investment in charging infrastructure. Under this assumption, the share of shifted EV load ranges between 77% and 80% compared to uncontrolled charging. When charging infrastructure costs are included and vary from high to low, the shifted EV load decreases to 44–59% in the V1G scenarios (Supplementary Figure 8).

Moreover, spiky charging profiles result in load factors of 1–20%, while per-vehicle charging capacity increases to 11.4–16.3 kW, an order of magnitude above the AFIR target. This result highlights that studies that neglect charging infrastructure costs may draw misleading conclusions, as the implied availability of charging power is substantially higher than when infrastructure costs are explicitly considered. As charging infrastructure costs increase, deployed capacity declines and utilization rates rise, reflecting a more streamlined deployment of costly assets. Conversely, when charging infrastructure costs decrease, EV scheduling becomes increasingly driven by electricity price signals.

Charging infrastructure costs also strongly affect the V2G operations. When costs are excluded, unidirectional and bidirectional charging infrastructure are assumed to have identical costs, leading to high V2G injection volumes. V2G energy injection increases from a maximum of 89 TWh in the Cost-Low, Limited Grid Expansion case to 378–389 TWh when charging infrastructure costs are null (Supplementary Figure 9). As a result, the maximum V2G revenues rise sharply from 6.4 billion € (Limited Grid Expansion, Cost-Low scenario) to 17.2 billion €, indicating that neglecting infrastructure costs can substantially overestimate the economic potential of V2G (Supplementary Figure 10).

# Supplementary Note 5: Limitations

Despite its validity, the model used to generate this study's results brings along a few limitations.

The spatial resolution of our model is national, with each country represented as a single node in the energy system. This choice reflects a trade-off between spatial and temporal resolution, with the latter deliberately set to one hour to capture hourly dispatch decisions and electricity price dynamics. A finer spatial granularity could reveal additional benefits from vehicle-to-grid (V2G) operation, particularly if a higher-resolution representation of the transmission network introduces additional congestion constraints. While this aspect is not explicitly addressed in this study, insights can be inferred from the scenarios exploring different levels of transmission grid expansion. In particular, lower interconnector capacities tend to increase the system value of V2G.

Furthermore, the distribution grid is not explicitly modeled, even though it could represent an additional bottleneck. Instead, potential distribution-level constraints are approximated by explicitly representing charging infrastructure, which is assigned a non-zero cost. The charging infrastructure load factor is endogenously optimized, reducing charging peaks even during periods of abundant, low-cost solar generation. This mechanism can be interpreted as a proxy for transformer or feeder limits at the distribution level. As a result, higher charging infrastructure costs impose stronger penalties on peak charging demand, effectively shaving charging peaks. While these constraints are not explicitly modeled, the introduction of charging infrastructure costs enables the model to capture some of the underlying dynamics associated with distribution grid limitations.

Moreover, this study does not consider V2G's potential to provide ancillary services to the power system, but only the effects arising from its interaction with the balancing electricity market.

In this study, we focus our analysis on private passenger vehicles, although other transport segments, including heavy-duty, light-duty, bus, and motorcycle segments, are represented in the model. This choice is primarily driven by the substantial uncertainty associated with generating realistic mobility profiles for non-passenger vehicle segments. In addition, the potential participation of these segments in smart charging or V2G remains highly uncertain, given their utilization patterns and operational requirements. Passenger vehicles, by contrast, are parked for at least 16 hours per day on average, with even longer parking durations during weekends when only inactive parking is considered (Pasaoglu et al., 2012).

The present study also does not account for the specific spatial location of charging infrastructure, which can influence both deployment decisions and utilization rates. Nevertheless, we estimate upper and lower bounds for the system-level savings achievable through charging infrastructure deployment, thereby providing a range within which its impact is expected to lie.

To limit model complexity and computational burden, the model does not include an explicit representation of gaseous fuel transport networks, and fuel transport is assumed to occur at no additional cost. Countries are therefore allowed to trade freely in fuels such as hydrogen, synthetic diesel, synthetic methane, synthetic methanol, and synthetic kerosene at no cost. Moreover, the European energy system is modeled in isolation, without considering imports from or exports to regions outside Europe. In the context of transport decarbonization, these assumptions tend to favor hydrogen-based technologies. Nevertheless, even under these optimistic conditions, the model results indicate that electric powertrains remain more cost-effective, particularly when smart charging strategies are enabled. These assumptions, therefore, provide a conservative test of the competitiveness of electric vehicles and further support the robustness of the study's conclusions.

The authors emphasize that the future adoption of V1G and V2G, along with their associated economic benefits, should not be interpreted as intrinsic country-specific characteristics. Instead, these outcomes depend primarily on the configuration of the underlying energy system, including generation mix, flexibility options, and transmission assumptions. The country-level results presented in this study, therefore, reflect conditional responses to the modeled energy system configurations rather than inherent national trends.

# Supplementary Note 6: Charging requirements and utilization rate

Here, we detail the computation of the Charging Requirement Coefficient (CRC, Equation 4). In the derivation of Equation 5, the load factor is defined as the ratio between the numerator and the denominator. The numerator, given by the sum of the energy flows handled by the charging infrastructure, can be decomposed into three components: the charging demand associated with vehicle mobility, the electricity injected into the grid through V2G operation, and the additional charging required to compensate for V2G discharge. The denominator represents the maximum energy flow that the charging infrastructure can accommodate.

$$\begin{matrix}\text{charging}\\\text{requirements}\end{matrix} = \frac{1}{\text{load factor}} \times \text{CRC} \tag{4}$$

$$\begin{aligned}
\text{CRC} &= \begin{matrix}\text{charging}\\\text{requirements}\end{matrix} \times \text{load factor} = \\
&= \frac{\text{peak power}_{(\text{V1G+V2G})}}{\text{vehicle number}} \times \frac{\text{charge}_{\text{mob}} + \text{charge}_{\text{V2G}} + \text{discharge}_{\text{V2G}}}{\text{peak power}_{(\text{V1G+V2G})} \times \text{timesteps}} = \\
&= \frac{1}{\frac{\text{charge}_{\text{mob}}}{\text{avg cons} \times \text{avg distance}}} \times \frac{\text{charge}_{\text{mob}} + \text{charge}_{\text{V2G}} + \text{discharge}_{\text{V2G}}}{\text{timesteps}} = \\
&= \frac{\text{avg cons} \times \text{avg distance}}{\text{charge}_{\text{mob}}} \times \frac{\text{charge}_{\text{mob}} \left(1 + \frac{\text{charge}_{\text{V2G}} + \text{discharge}_{\text{V2G}}}{\text{charge}_{\text{mob}}}\right)}{\text{timesteps}} = \\
&= \frac{\text{avg cons} \times \text{avg distance}}{\text{timesteps}} \times \left(1 + \frac{\text{charge}_{\text{V2G}} + \text{discharge}_{\text{V2G}}}{\text{charge}_{\text{mob}}}\right) = \\
&= \frac{\text{avg cons} \times \text{avg distance}}{\text{timesteps}} \times \left(1 + \frac{\text{discharge}_{\text{V2G}} \times \left(1 + \frac{1}{\eta_{V2G}}\right)}{\text{charge}_{\text{mob}}}\right)
\end{aligned} \tag{5}$$

The CRC is directly proportional to the average vehicle energy consumption ($kWh\ vehicle^{-1}$), the annual average traveled distance (km vehicle$^{-1}$ year$^{-1}$), the time horizon in hours, and a term accounting for electricity discharged through V2G operations. Specifically, the last term represents the ratio between the electricity discharged through V2G operations and the additional charging required to meet the mobility demand of the EV fleet, where $\eta_{\text{V2G}}$ denotes the round-trip efficiency of the V2G chargers. The sequence of calculations used to derive equation (2) is provided in the Supplementary Information.

The same CRC value yields the same curves when the charging requirements are plotted against the load factor. Higher average consumption, or higher average distance per vehicle, shifts the curve upward, increasing charging requirements. Likewise, V2G operations produce the same results of shifting the curve upward. Supplementary Figure 11 shows the results for the Cost-Base case.

# Supplementary Note 7: Calculation of charging requirements upper and lower bounds

We define lower- and upper-bound charging requirements. The lower bound corresponds to the charging deployment resulting from the energy system optimization, while the upper bound is defined as the maximum charging deployment that yields non-negative system benefits. The computation of the upper bound follows three main steps, which are illustrated in Supplementary Figure 12.

(a) We first isolate the energy system benefits in monetary terms. We sum the energy system benefits across all countries, noting that some countries may exhibit negative values. To account for this, negative system benefits are offset by redistributing the total monetary benefits only to countries with positive system benefits.

(b) The resulting monetary system benefits are then divided by the average charging infrastructure cost, including both investment and operation and maintenance costs. This average cost is computed as a weighted mean of unidirectional and bidirectional charging infrastructure costs, based on their respective deployment levels. This allows us to estimate the additional charging infrastructure capacity that could be deployed. By preserving the same capacity-weighted cost structure, we implicitly assume that the expansion maintains the same bidirectional charging share in V2G scenarios. The total charging capacity is then updated by adding the newly deployable capacity.

(c) Starting from the augmented charging capacity, we compute the corresponding charging requirements, assuming that the number of vehicles remains constant. The Charging Requirement Coefficient of the upper bound is set equal to that of the corresponding lower-bound scenario, as electric vehicle consumption parameters are unchanged. This allows us to derive the associated load factor and to position the upper-bound point in the charging requirement versus load factor space.

# Supplementary Note 8: Vehicle cost assumptions

We included the cost of vehicles in the energy system model. Each transport segment and powertrain is modeled with its investment cost and, when available, with its variable and fixed costs. The values for passenger vehicles and light-duty vehicles are retrieved from Burke et al. (2024), for heavy-duty vehicles and buses from the Danish Energy Agency (L1 and B1 types) (Danish Energy Agency, 2024), while for motorcycles, we applied our own assumptions. All the cost assumptions are reported in Table 2. Vehicle costs are then normalized to the peak transport demand to account for vehicle inutilization.

# Supplementary Note 9: Charging infrastructure

We define four charging infrastructure cost levels, including a zero-cost case that serves as a reference for comparison. The cost assumptions are based on Herbst et al. (2023). Although a broader review of available estimates was conducted, the results of which are summarized in Supplementary Figure 13. Charging infrastructure costs are difficult to compare across studies because cost breakdowns are often inconsistent, and it is not always clear whether the underlying charging station layouts, power ratings, or utilization assumptions are comparable. As a result, real-world costs may differ from reported values.

In addition to the baseline cost level, we consider two alternative cost assumptions to perform a sensitivity analysis. These alternative levels reflect both potential cost reductions over time and differences across charging infrastructure types. In particular, charging costs per unit of installed capacity can vary significantly depending on station layout, power level, and technological configuration. Since the present study does not explicitly model the spatial location of chargers or the composition of charging technologies, multiple cost levels are introduced to capture this uncertainty and to assess its implications for system outcomes. Selected values are reported in Table 3.

For reference, in 2025, BMW launched its V2G offer in collaboration with E.ON. In Germany, customers can choose between the unidirectional BMW Wallbox Plus at around 699 € and the bidirectional BMW Wallbox Professional at about 2095 €, which is required for V2G with the iX3 and a special E.ON tariff. Under the tariff, participants can receive around 24 cents per hour for every hour the car is connected, capped at about 720 € per year, but it remains unclear how much of the V2G-capable charging infrastructure is subsidized (Schwarzer, 2025).

The analysis does not attempt to project future charging equipment and installation costs. As reported in (Wood et al., 2023), stakeholder interviews suggested that future cost trends are highly uncertain, with potential cost reductions from economies of scale and learning effects offset by supply chain constraints and site-specific installation challenges.

# Supplementary Figures

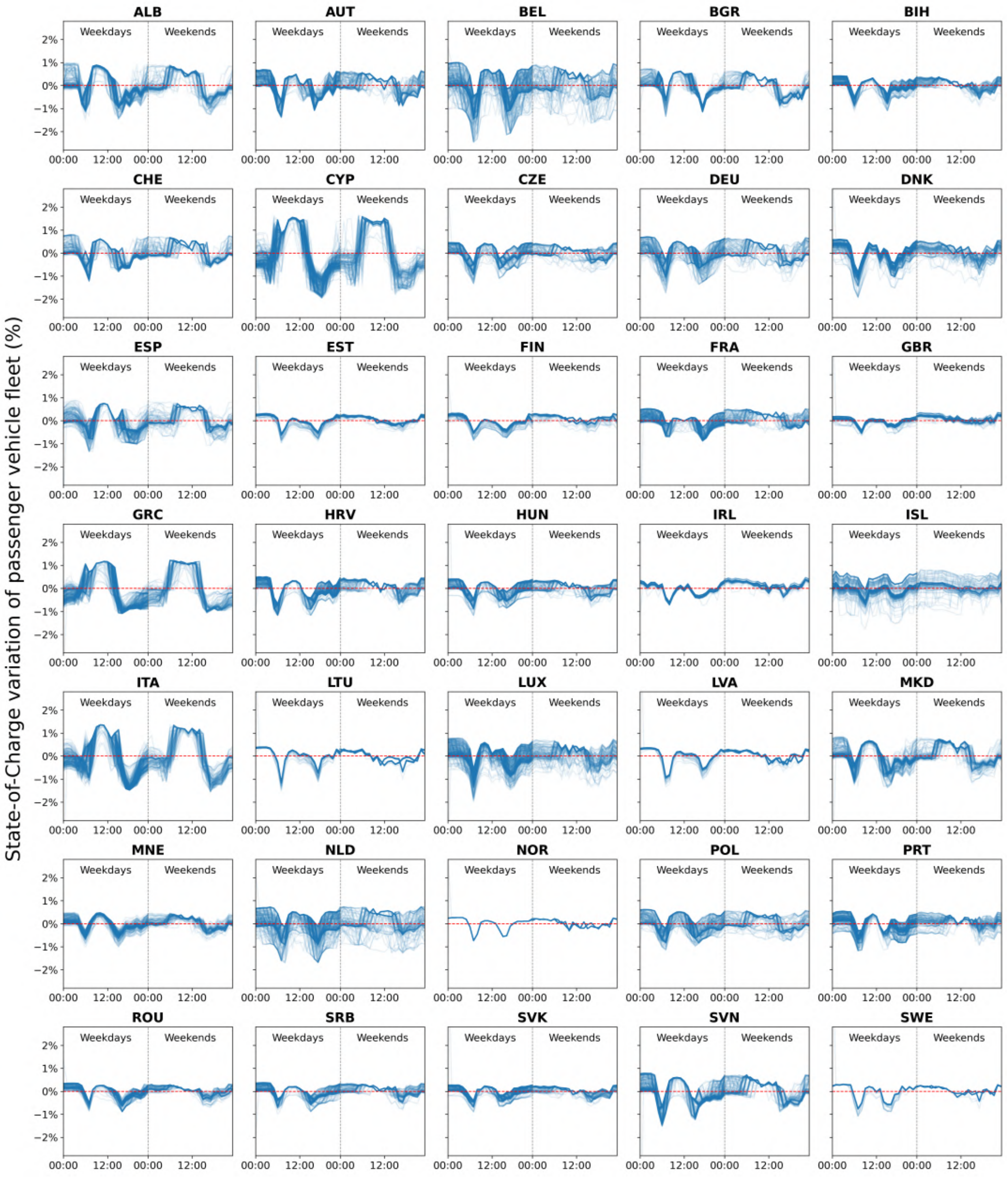

**Supplementary Figure 1: Hourly State-of-Charge (SoC) variation of the aggregated EV passenger fleet in European countries.** The SoC time series are overlaid by weekday and weekend categories. The figure shows results for the V2G charging scenario under the Limited Grid Expansion and Cost-Low cases, which represent conditions in which V2G is most intensively used compared to other non-zero charging infrastructure cost scenarios. The hourly SoC variation does not exceed 2%, indicating that EV fleet batteries are not overused.

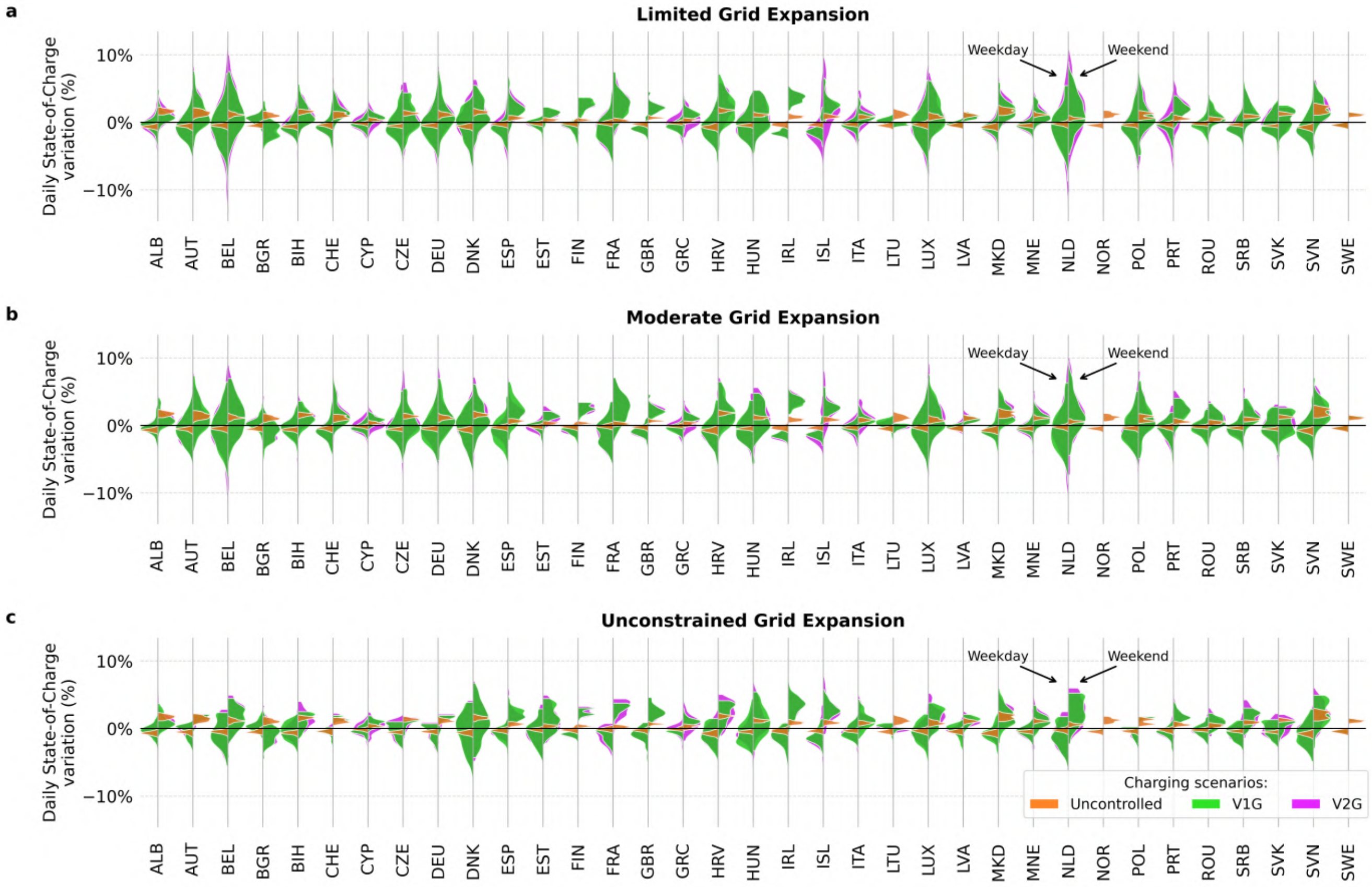

**Supplementary Figure 2: Daily State-of-Charge (SoC) variation on weekdays and weekends.** The figure shows, for each modelled country, a violin plot in which each half represents the cumulative SoC differences for weekdays and weekends. The results refer to the Moderate Grid Expansion scenario in the Cost-Low case.

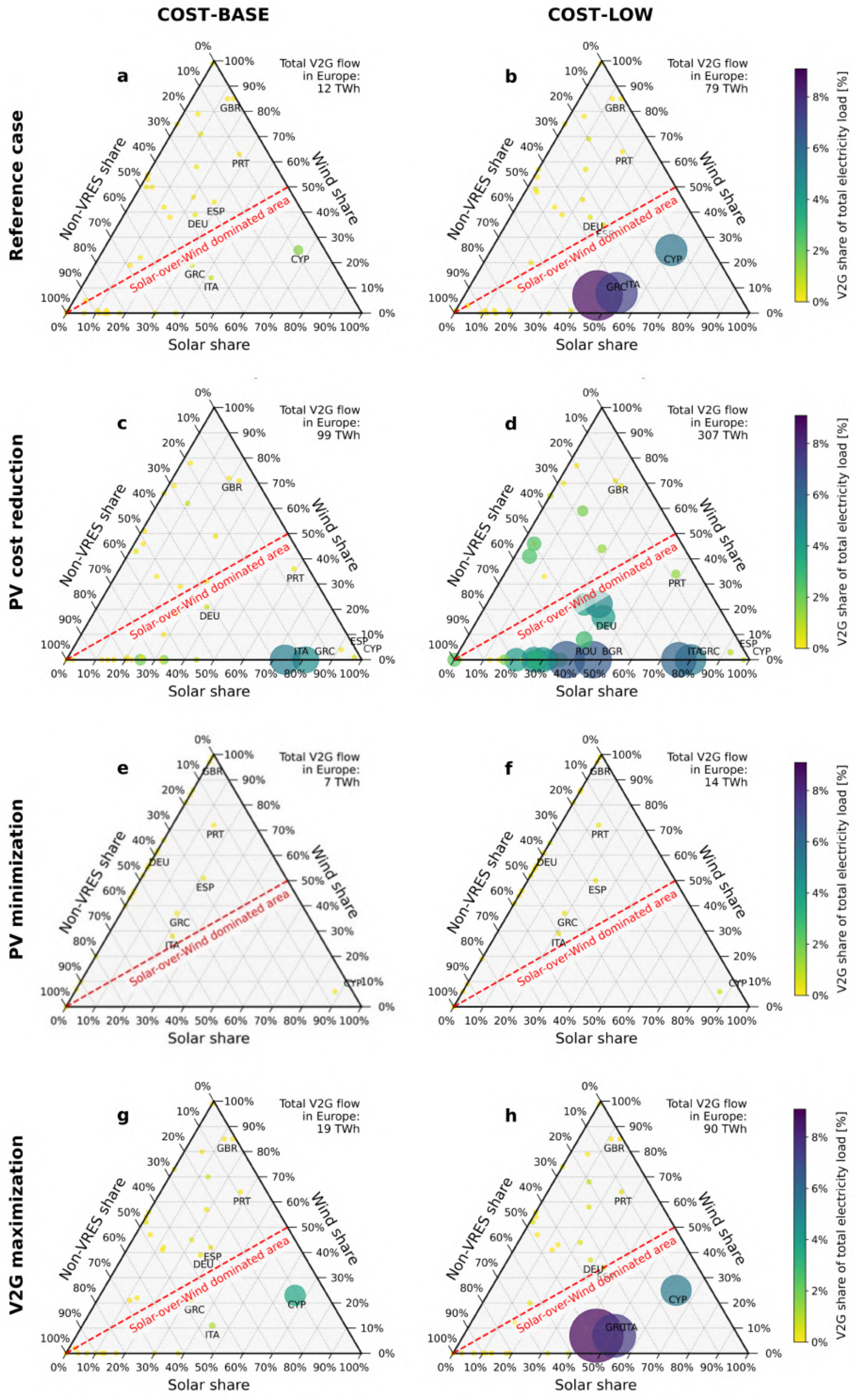

**Supplementary Figure 3: Ternary plots of sensitivity analysis.** The figure presents ternary plots of energy system configurations obtained from the sensitivity analysis for two cases, Cost-Base and Cost-Low, with V2G enabled under the Moderate Grid Expansion scenario. (a–b) reference case; (c–d) reduced open-field PV investment costs by 50%; (e–f) target MGA analysis minimizing the deployment of PV technologies; (g–h) target MGA analysis maximizing the utilization of V2G. In both MGA analyses, a 10% slack cost is applied to the total system cost, including vehicle costs.

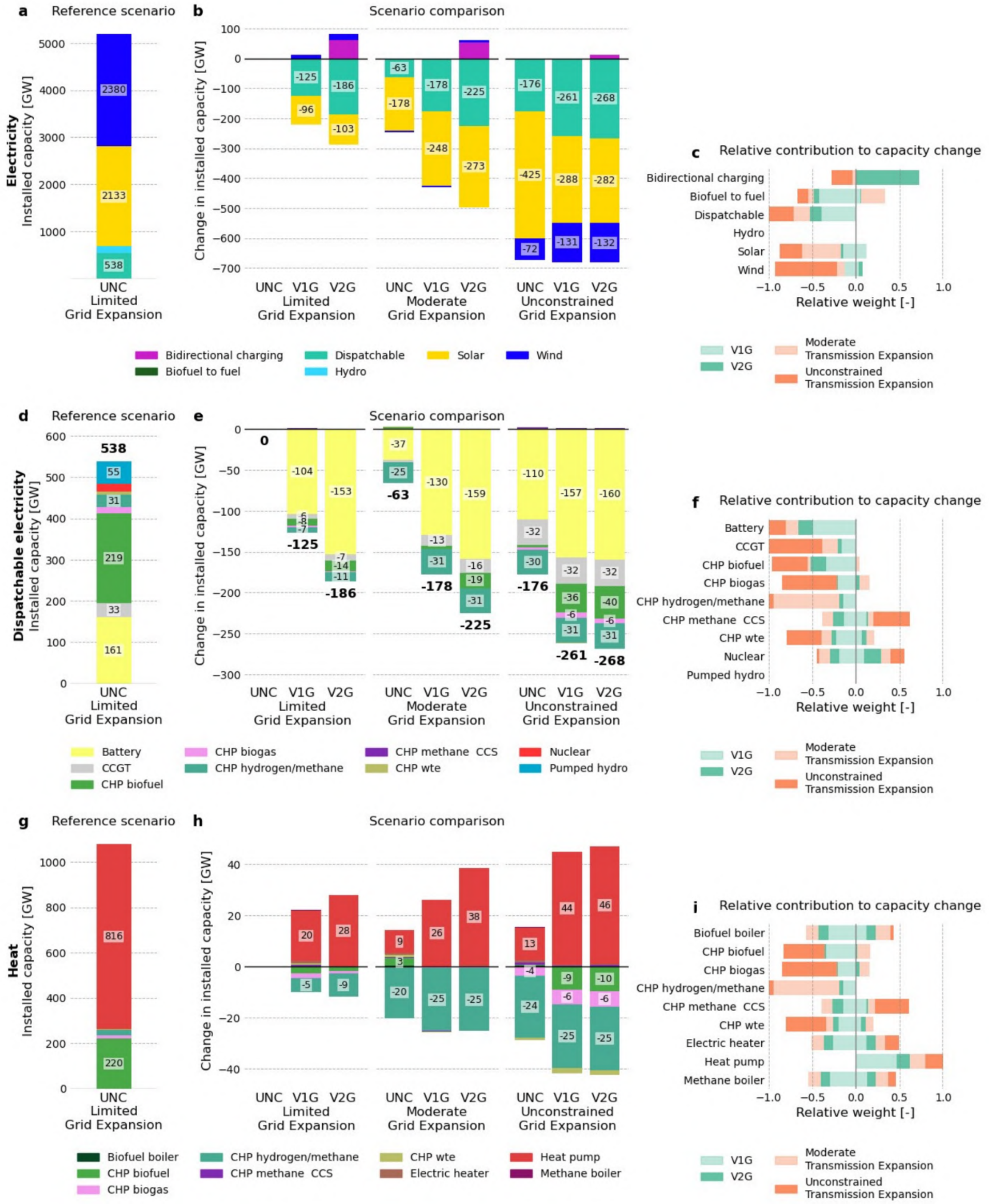

**Supplementary Figure 4: Capacity deployment sensitivity check using Shapley decomposition analysis.** The figure shows the changes in capacity deployment for the Cost-Base scenarios. (a,d,g) Uncontrolled charging scenarios, reporting the composition of the capacity mix for each sector or sub-sector; (b,e,h) relative changes in capacity deployment compared to the reference scenario; (c,f,i) Shapley decomposition evaluates the main drivers of capacity changes.

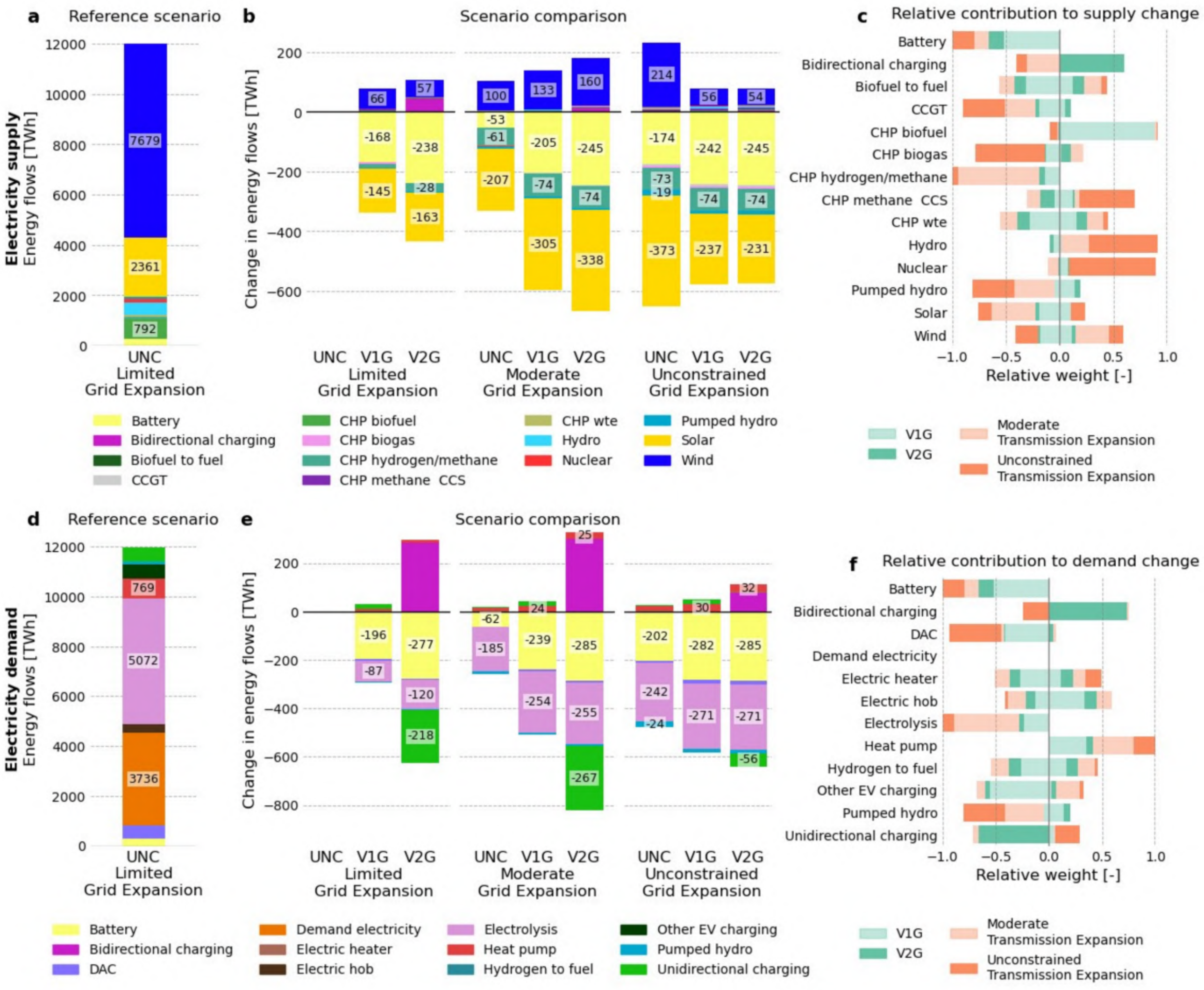

**Supplementary Figure 5: Supply and demand flow sensitivity check using Shapley decomposition analysis.** The figure shows the changes in the supply and demand of electricity flows for the Cost-Base scenarios. (a,d,g) Uncontrolled charging scenarios, reporting the composition of the supply and demand mix for the electricity sector; (b,e,h) relative changes in flows compared to the reference scenario; (c,f,i) Shapley decomposition evaluates the main drivers of capacity changes.

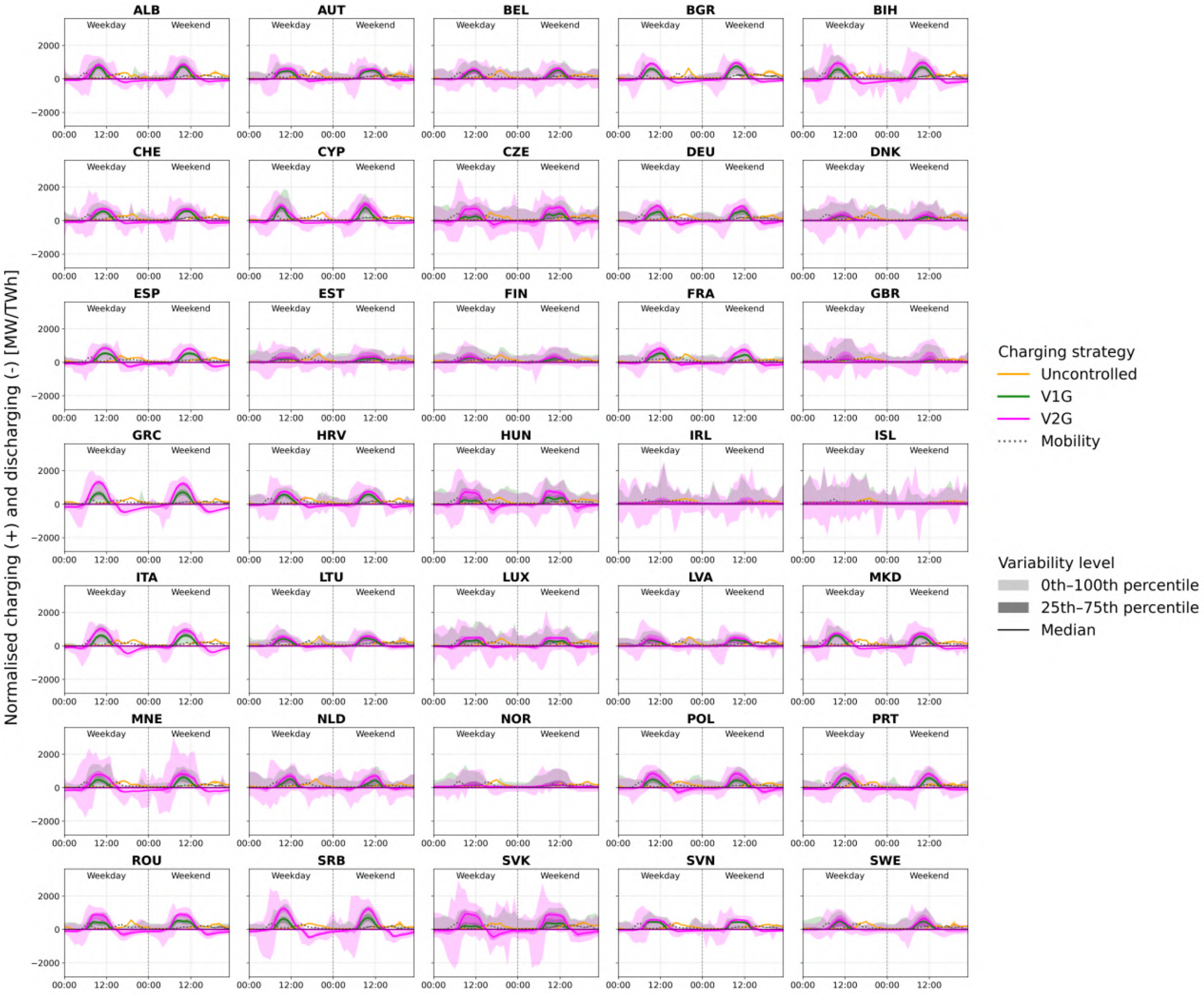

**Supplementary Figure 6: Hourly charging and mobility time series in the Cost-Zero, Moderate Grid Expansion scenario.** The shaded areas represent the different percentiles of values for each charging option (Uncontrolled, V1G, and V2G), displayed in different colors. The profiles are aggregated according to weekday and weekend day.

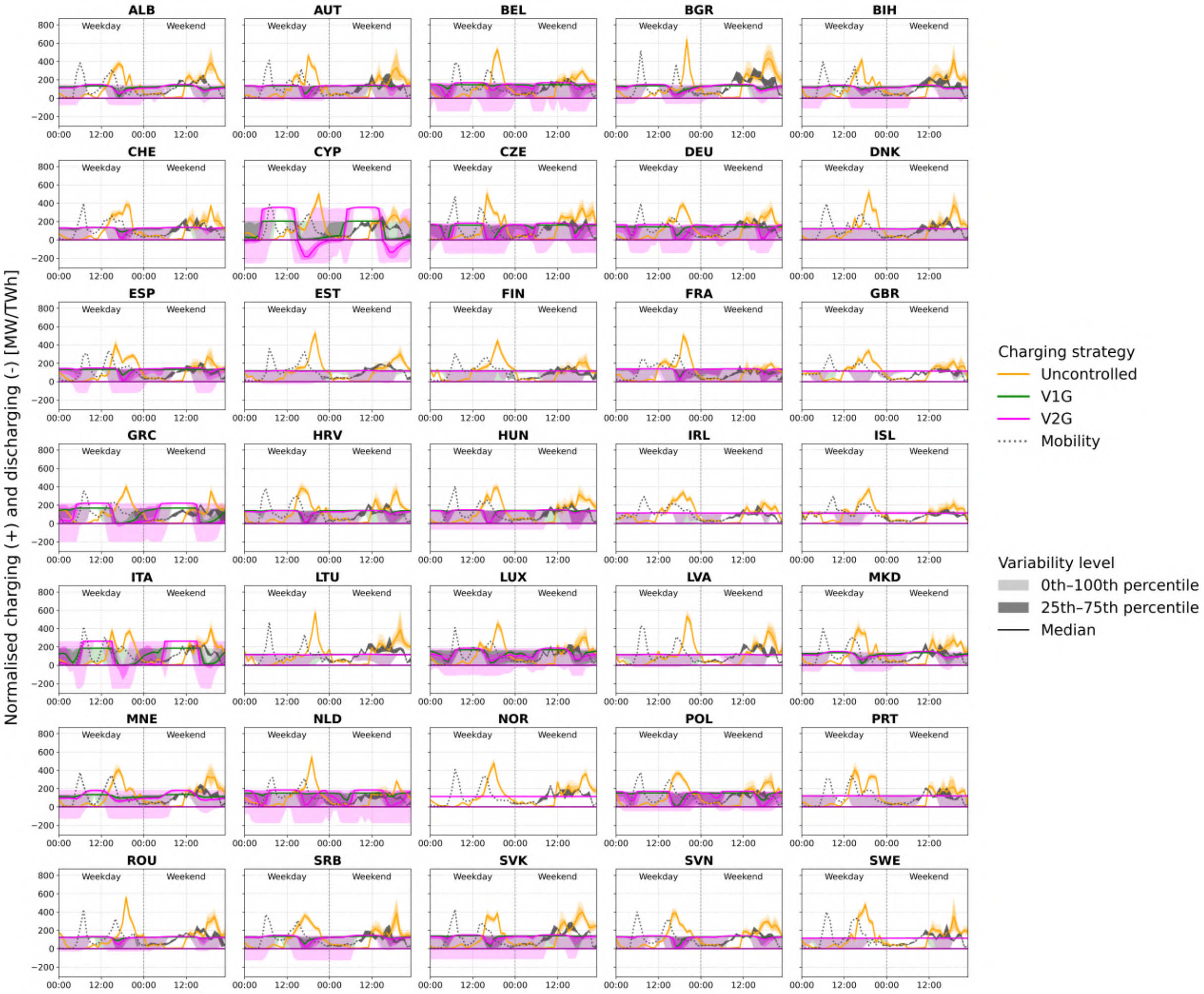

**Supplementary Figure 7: Hourly charging and mobility time series in the Cost-Base, Moderate Grid Expansion scenario.** The shaded areas represent the different percentiles of values for each charging option (Uncontrolled, V1G, and V2G), displayed in different colors. The profiles are aggregated according to weekday and weekend day.

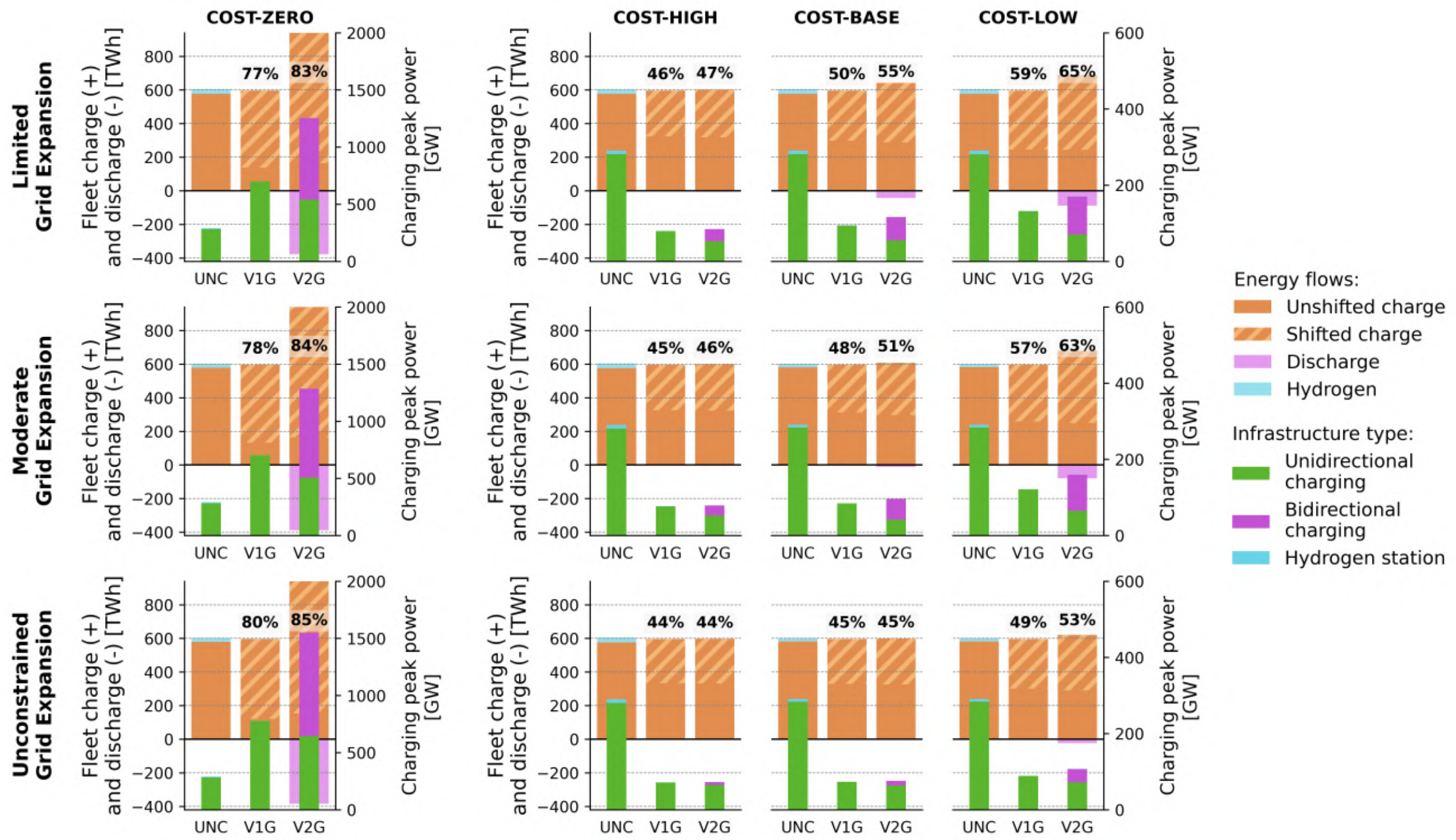

**Supplementary Figure 8: Fleet charge and discharge and charging peak power capacity.** The figure displays on the left-end axis the charging and discharging flows relative to the passenger fleets, including all types of powertrains. The shaded area represents the amount of charging shifted relative to the uncontrolled case. On the right axis, the figure shows the split of the total charging power peak by charging infrastructure type: unidirectional or bidirectional. The annotations refer to the amount of shifted electricity for charging EVs.

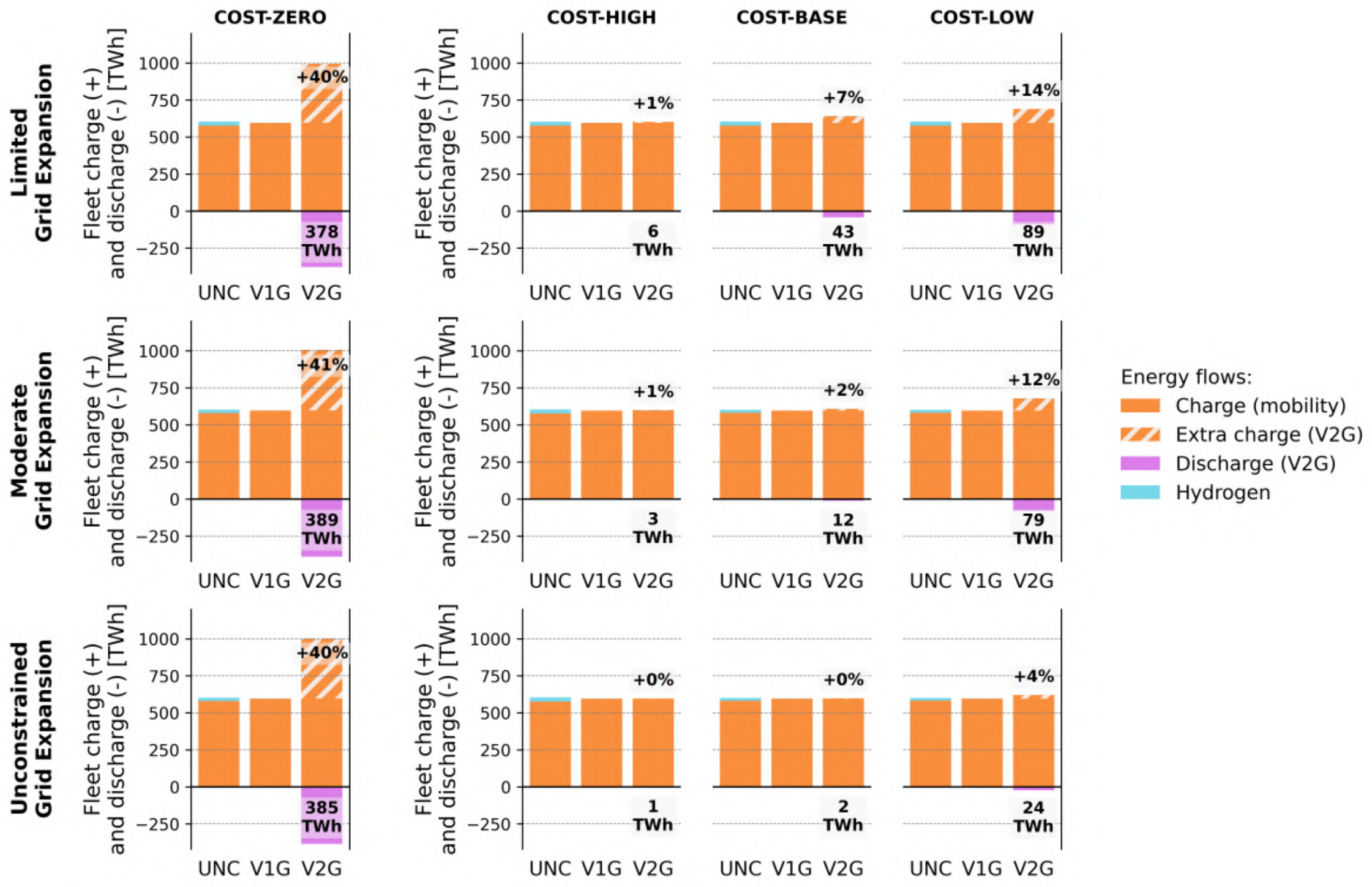

**Supplementary Figure 9: Fleet charge and discharge, and charging needs to support V2g operations.** The figure displays on the left-end axis the charging and discharging flows relative to the passenger fleets, including all types of powertrains. The shaded area represents the additional charging required, in addition to the electricity used for mobility, to support V2G discharge in later stages. The annotations refer to the amount of charge electricity necessary to support V2G discharge in addition to the amount required to satisfy the mobility demand.

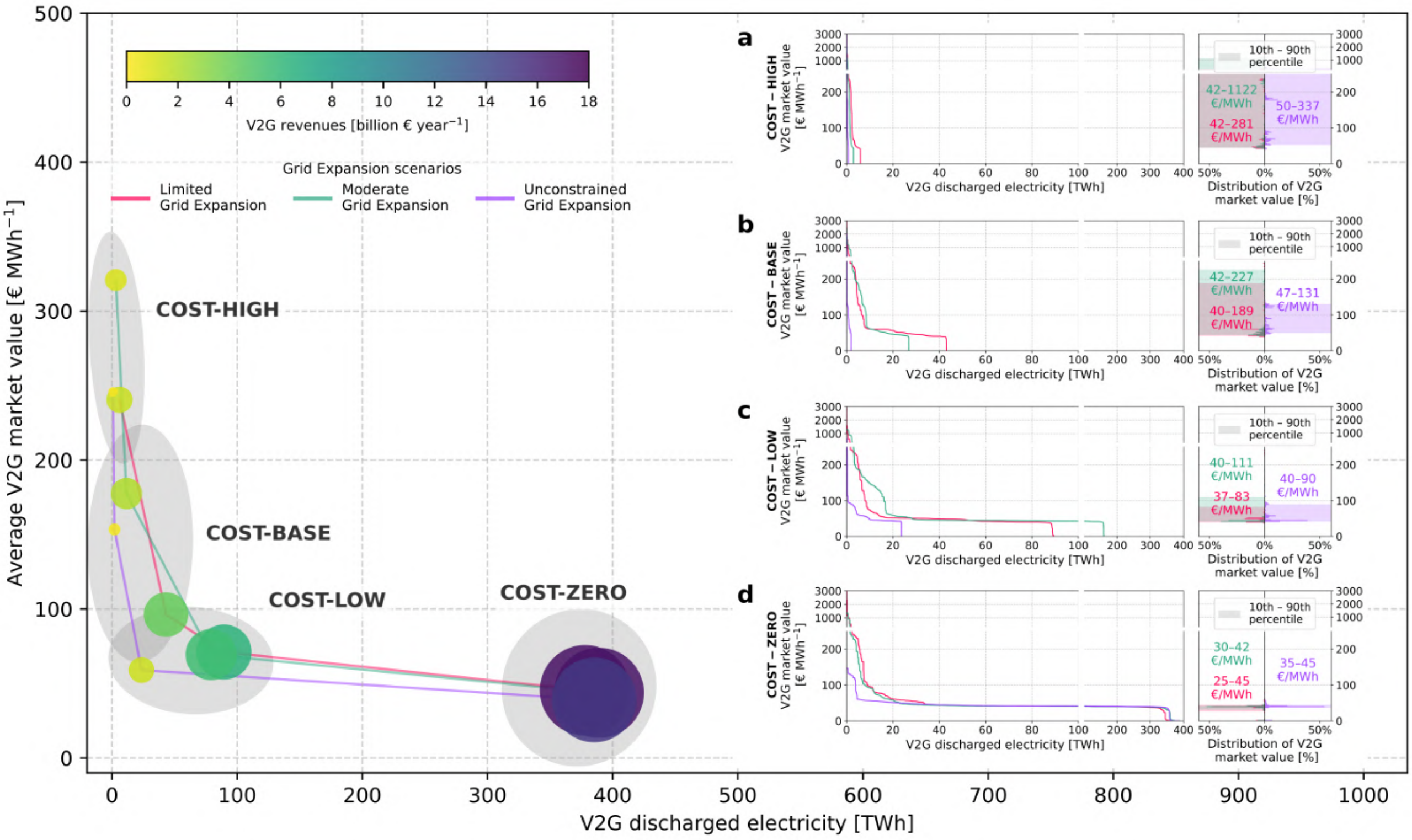

**Supplementary Figure 10: V2G revenues and market remuneration.** The main figure shows the potential revenues from V2G deployment as a function of discharged electricity (injection into the grid) on the x-axis and the average V2G market value on the y-axis. The size and colour of the circles represent the total revenues from V2G operations, computed using the shadow price corresponding to the timesteps at which electricity is discharged from the battery and injected into the grid. The coloured lines connect the same Grid Expansion scenarios across different charging infrastructure cost scenarios. Panels (a–d) show two subplots: on the left, the V2G market value as a function of the hourly injected quantity for each country; on the right, the weighted distribution of V2G market values based on V2G injection volumes.

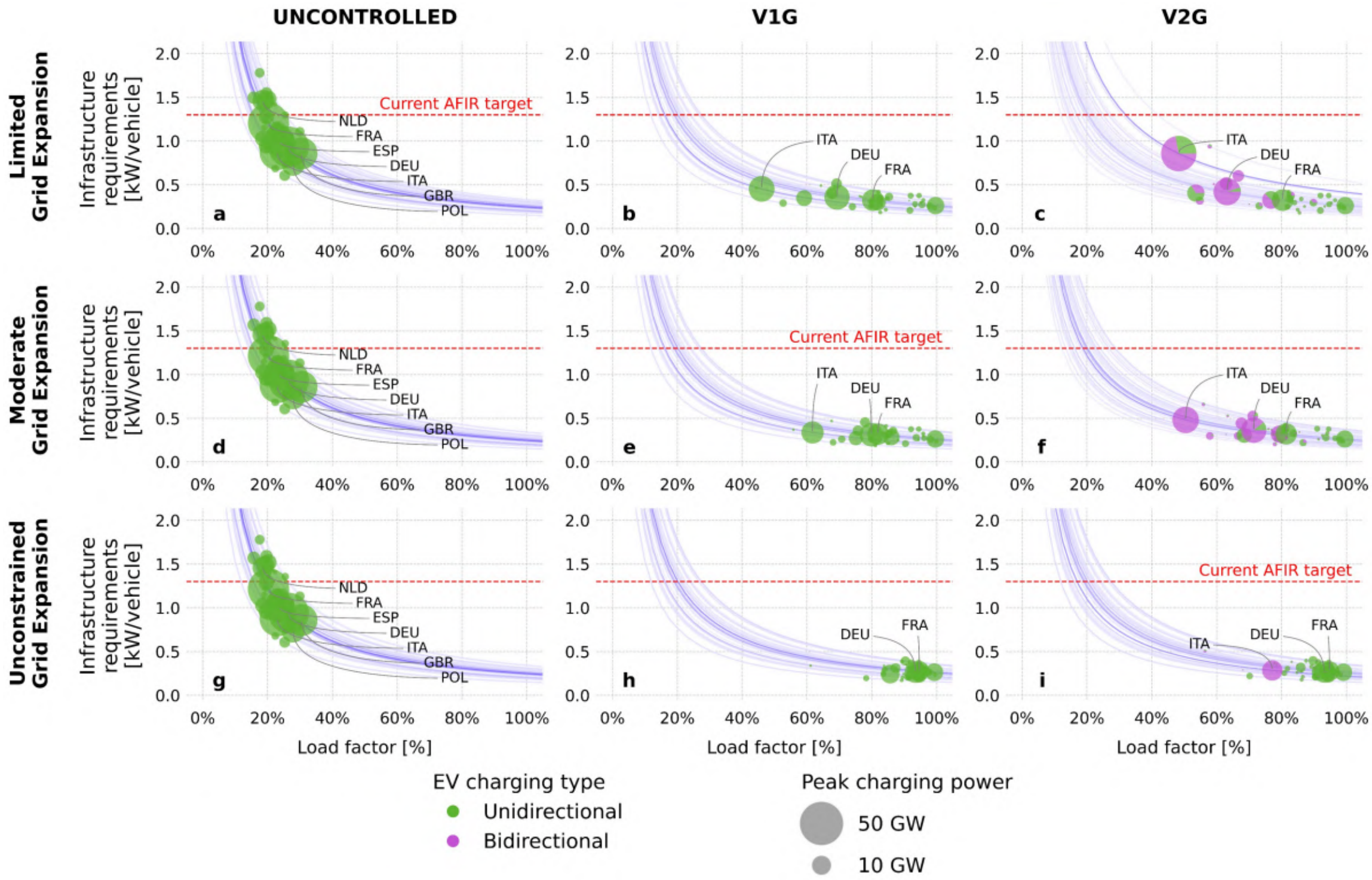

**Supplementary Figure 11: Charging infrastructure requirements versus load factor.** The figure shows the relationship between charging infrastructure requirements for passenger EV fleets across different countries, represented by pie charts. Each pie wedge indicates the type of charging infrastructure, distinguishing between unidirectional (green) and bidirectional (fuchsia) charging. The blue curves represent the theoretical relationship between charging infrastructure requirements and load factors, which depend on per-vehicle charging demand for mobility purposes and on V2G injections. Countries with the same per-vehicle mobility consumption share the same curve. When V2G is enabled, the curve shifts upward and to the right, allowing higher load factors while keeping charging infrastructure requirements constant. Charging infrastructure requirements are computed based on each country's peak charging demand and number of vehicles.

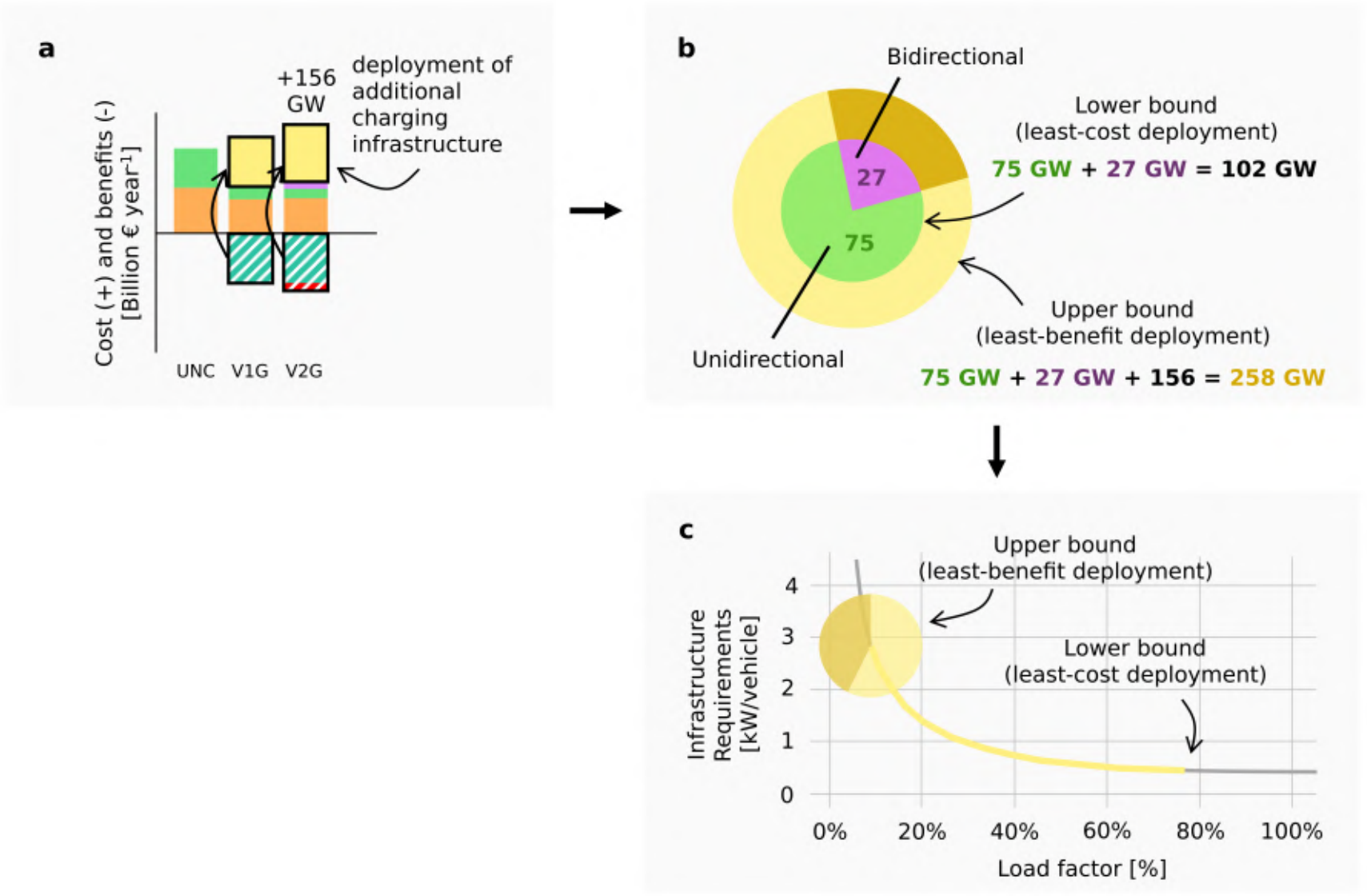

**Supplementary Figure 12: Methodology for computing the maximum cost-neutral deployment of charging infrastructure.** The figure illustrates the steps used to compute the upper bound of charging infrastructure deployment, referred to as the *maximum cost-neutral deployment*. This represents the maximum level of charging infrastructure deployment—assuming the same mix of unidirectional and bidirectional chargers—that yields no additional total system cost savings compared to the Uncontrolled scenario. (a) Computation of system benefits (or cost savings) of the V1G and V2G scenarios relative to the Uncontrolled scenario. Countries with negative savings (i.e., additional costs) are set to zero, and total system cost savings are redistributed proportionally based on the initial cost-saving values. (b) Computation of the equivalent additional charging infrastructure that can be deployed while maintaining the same share of unidirectional and bidirectional charging infrastructure. The lower bound represents the cost-optimal result obtained from the energy system model. (c) Computation of charging infrastructure requirements based on the upper bound (maximum cost-neutral deployment), followed by interpolation on the corresponding country-specific curve.

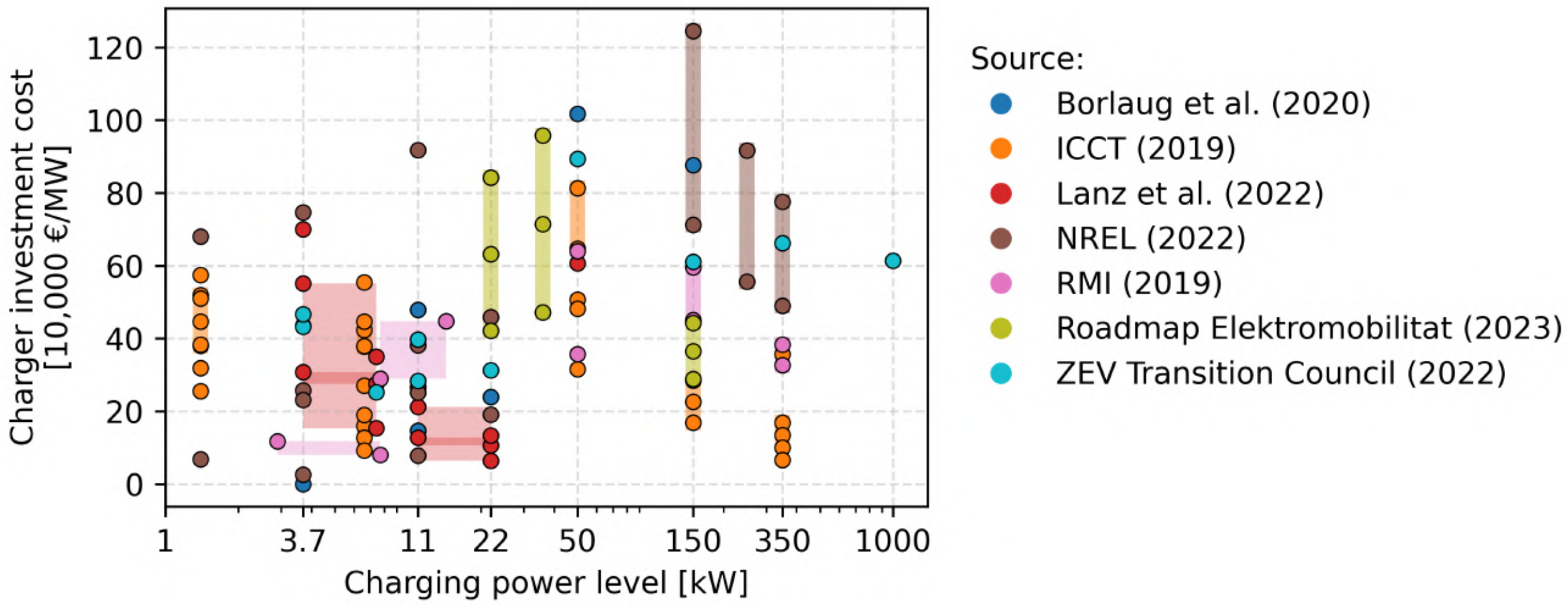

**Supplementary Figure 13: Charging infrastructure investment cost estimates.**

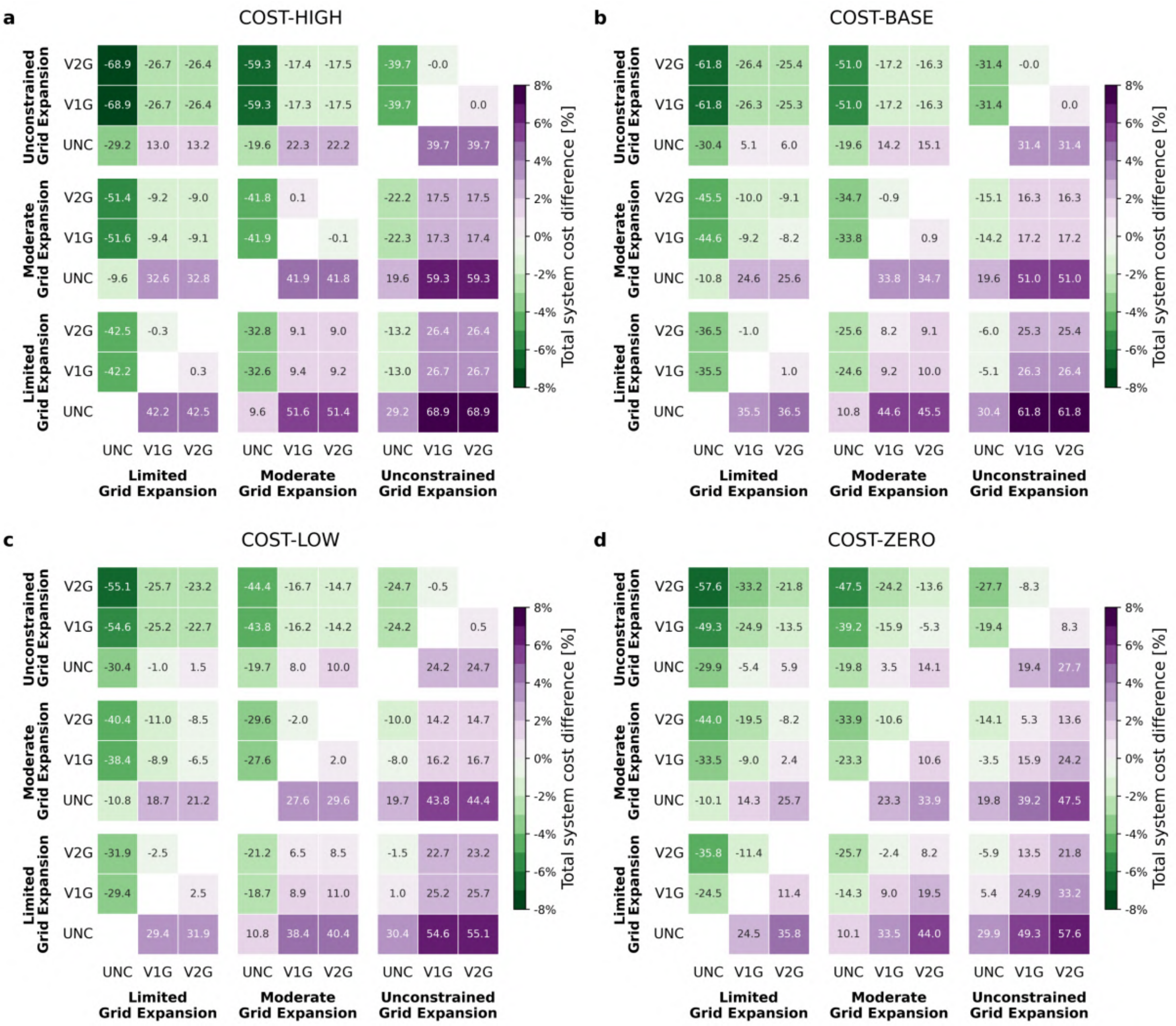

**Supplementary Figure 14: Total system cost savings.** Values represent the absolute energy system cost savings, expressed in billion €year$^{-1}$. The total system cost differences expressed in percentage exclude vehicle investment and operational costs, in order to ensure comparability with previous studies that did not account for vehicle costs.

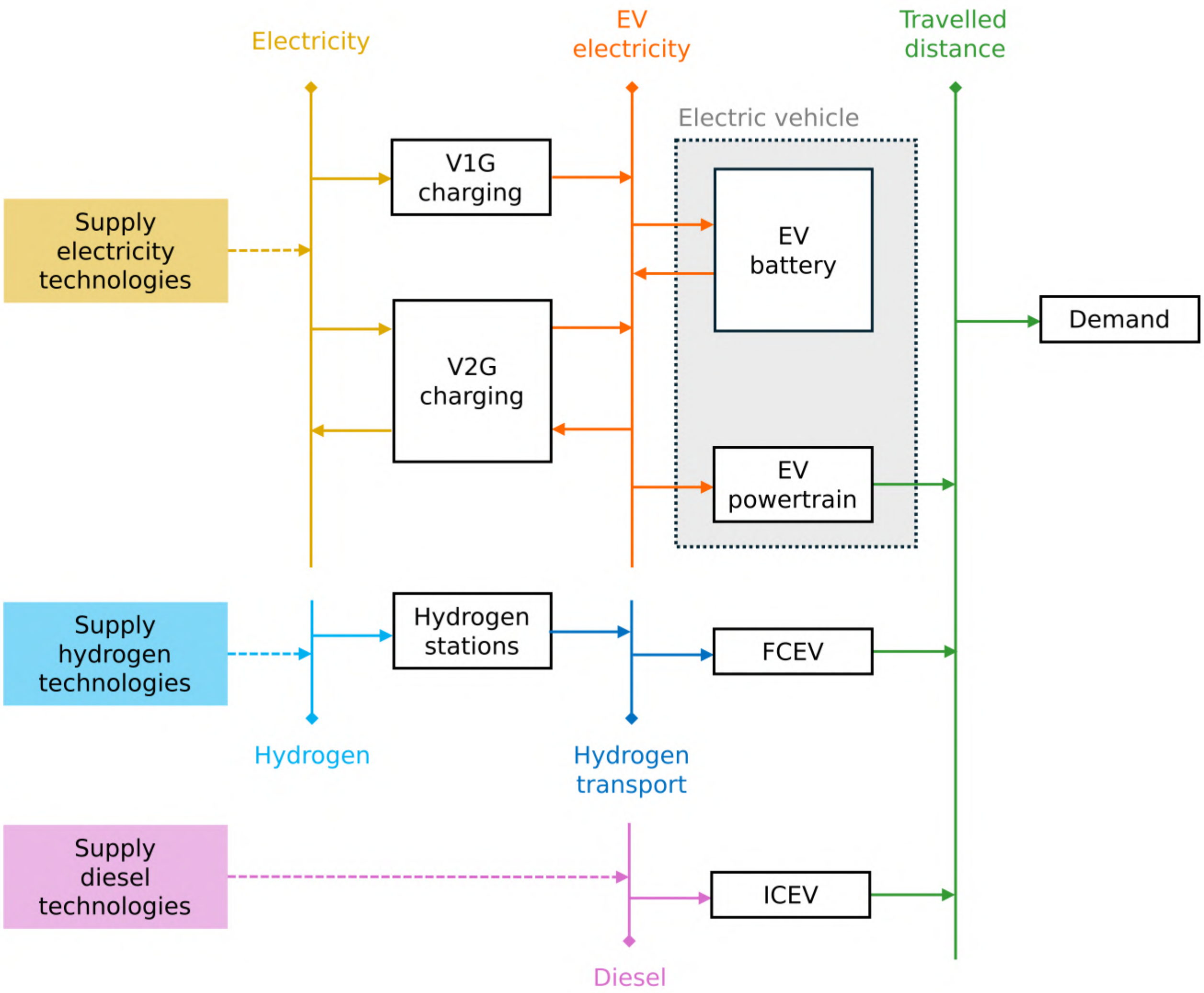

**Supplementary Figure 15: Schematic of the EV and charging infrastructure representation in the energy system model.**

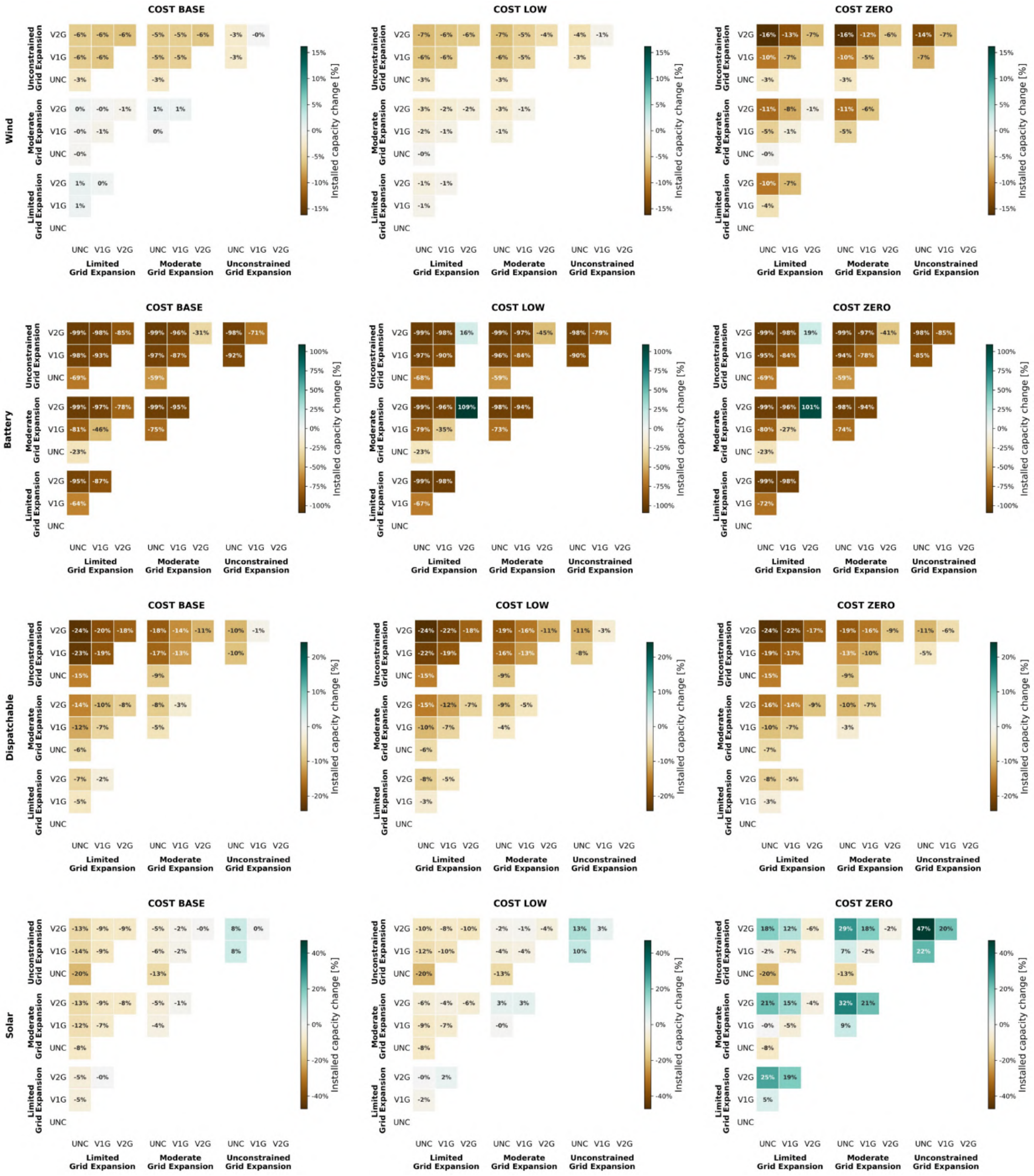

**Supplementary Figure 16: Relative installed capacity change.** The figure shows the relative change in deployed capacity with respect to the reference scenario, as defined by the rows and columns of the double-entry heatmap. The sign is consistent with the change occurring when moving from the scenarios on the x-axis to those on the y-axis.

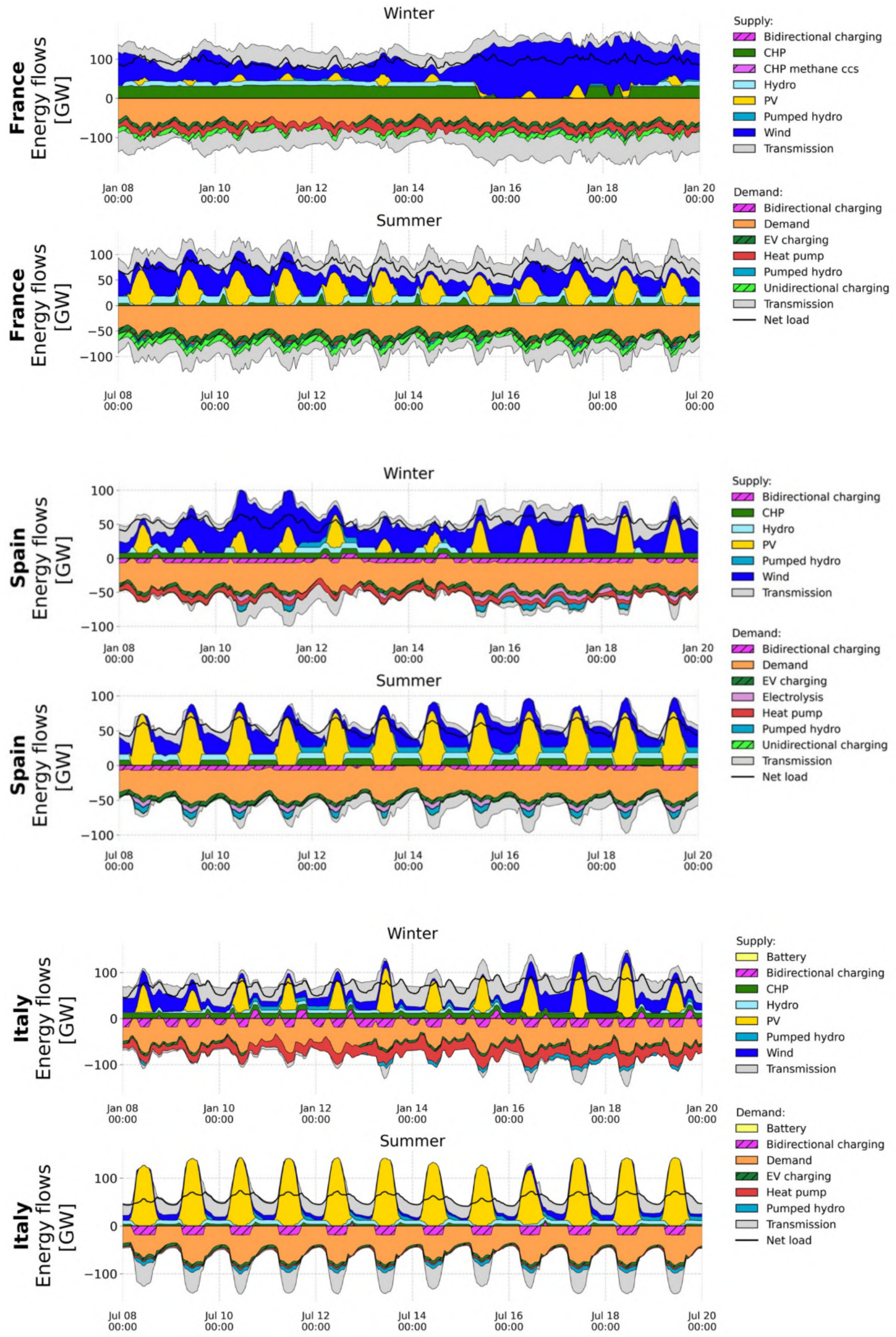

**Supplementary Figure 17: Country hourly dispatch in three European countries in the V2G — Cost-Base — Moderate Grid Expansion scenario.**

# Supplementary Tables

**Supplementary Table 1:** Sensitivity analysis on the V1G and V2G charging infrastructure costs. Tested scenario: V2G-Moderate Grid Expansion.

| Charging infrastructure cost scenario | Unidirectional charging (GW) | Bidirectional charging (GW) | Total charging (GW) | V2G share [%] | V2G discharge [TWh] |
|---|---|---|---|---|---|
| Cost-Base | 82 | 109 | 191 | 57 | 12 |
| Cost-Base (V1G cost = V2G cost) | 9 | 192 | 200 | 96 | 15 |
| Cost-Base (-50% PV cost) | 97 | 65 | 162 | 40 | 99 |
| Cost-Low | 129 | 189 | 318 | 60 | 79 |
| Cost-Low (V1G cost = V2G cost) | 21 | 305 | 325 | 94 | 82 |
| Cost-Low (-50% PV cost) | 139 | 159 | 298 | 53 | 307 |

**Supplementary Table 2:** Transport investment and operation costs in 2050. The values for passenger vehicles and light-duty vehicles are retrieved from Burke et al. (2024), for heavy-duty vehicles and buses from the Danish Energy Agency (L1 and B1 types) (Danish Energy Agency, 2024), while for motorcycles, we applied our own assumptions.

| **Vehicle type** | **Parameter** | **Unit** | **FCEV** | **EV** | **ICEV** |
|---|---|---|---|---|---|
| Passenger cars | Investment | EUR2015/vehicle | 20210 | 19832 | 18232 |
| | O&M variable | EUR2015/km | 0.021 | 0.008 | 0.027 |
| | O&M fixed | EUR2015/vehicle/year | 0.0 | 0.0 | 0.0 |
| | Lifetime | years | 12 | 12 | 12 |
| | Consumption | kWh/km | 0.252 | 0.180 | 0.375 |
| Light-duty vehicles | Investment | EUR2015/vehicle | 27360 | 24249 | 24424 |
| | O&M variable | EUR2015/km | 0.021 | 0.008 | 0.027 |
| | O&M fixed | EUR2015/vehicle/year | 0.0 | 0.0 | 0.0 |
| | Lifetime | years | 12 | 12 | 12 |
| | Consumption | kWh/km | 0.378 | 0.270 | 0.481 |
| Buses | Investment | EUR2015/vehicle | 201632 | 201632 | 159507 |
| | O&M variable | EUR2015/km | 0.1036 | 0.1036 | 0.1130 |
| | O&M fixed | EUR2015/vehicle/year | 6026.24 | 6026.24 | 6779.52 |
| | Lifetime | years | 13 | 13 | 13 |
| | Consumption | kWh/km | 2.266 | 1.620 | 3.021 |
| Motorcycles | Investment | EUR2015/vehicle | - | 12900 | 15000 |
| | O&M variable | EUR2015/km | - | 0.0 | 0.0 |
| | O&M fixed | EUR2015/vehicle/year | - | 0.0 | 0.0 |
| | Lifetime | years | - | 10 | 10 |
| | Consumption | kWh/km | - | 0.072 | 0.151 |
| Heavy-duty vehicles | Investment | EUR2015/vehicle | 218343 | 188465 | 199468 |
| | O&M variable | EUR2015/km | 0.1008 | 0.1008 | 0.1130 |
| | O&M fixed | EUR2015/vehicle/year | 6026 | 6026 | 6780 |
| | Lifetime | years | 14 | 14 | 14 |
| | Consumption | kWh/km | 2.517 | 1.800 | 2.849 |

**Supplementary Table 3:** Charging infrastructure cost assumptions by sensitivity case and type of charger.

| Cost scenario | Infrastructure | Cost type | Unit | Value |
|---|---|---|---|---|
| High | Unidirectional | Investment | 10'000 $EUR_{2015}$/MW | 84.273 |
| | | Fixed O&M | 10'000 $EUR_{2015}$/MW/yr | 1.873 |
| | Bidirectional | Investment | 10'000 $EUR_{2015}$/MW | 94.795 |
| | | Fixed O&M | 10'000 $EUR_{2015}$/MW/yr | 1.873 |
| Base | Unidirectional | Investment | 10'000 $EUR_{2015}$/MW | 63.205 |
| | | Fixed O&M | 10'000 $EUR_{2015}$/MW/yr | 1.346 |
| | Bidirectional | Investment | 10'000 $EUR_{2015}$/MW | 69.518 |
| | | Fixed O&M | 10'000 $EUR_{2015}$/MW/yr | 1.346 |
| Low | Unidirectional | Investment | 10'000 $EUR_{2015}$/MW | 42.136 |
| | | Fixed O&M | 10'000 $EUR_{2015}$/MW/yr | 0.819 |
| | Bidirectional | Investment | 10'000 $EUR_{2015}$/MW | 44.241 |
| | | Fixed O&M | 10'000 $EUR_{2015}$/MW/yr | 0.819 |
| Zero | Unidirectional | Investment | 10'000 $EUR_{2015}$/MW | 0.0 |
| | | Fixed O&M | 10'000 $EUR_{2015}$/MW/yr | 0.0 |
| | Bidirectional | Investment | 10'000 $EUR_{2015}$/MW | 0.0 |
| | | Fixed O&M | 10'000 $EUR_{2015}$/MW/yr | 0.0 |

**Supplementary Table 4:** Estimated battery capacity per country and transport segment in 2050 (TWh). We assume battery capacities of 80 kWh for passenger vehicles, 150 kWh for light-duty vehicles, 15 kWh for motorcycles, 500 kWh for buses, and 1000 kWh for heavy-duty vehicles.

| Country | Light-duty | Heavy-duty | Passenger car | Motorcycle | Bus | Total |
|---|---|---|---|---|---|---|
| AUT | 0.0634 | 0.0536 | 0.3983 | 0.0080 | 0.0050 | 0.5284 |
| ALB | 0.0090 | 0.0109 | 0.0368 | 0.0004 | 0.0036 | 0.0606 |
| BEL | 0.1139 | 0.0967 | 0.4683 | 0.0074 | 0.0081 | 0.6943 |
| BGR | 0.0523 | 0.0385 | 0.2219 | 0.0017 | 0.0104 | 0.3248 |
| CHE | 0.0562 | 0.0422 | 0.3733 | 0.0111 | 0.0072 | 0.4900 |
| CYP | 0.0148 | 0.0105 | 0.0441 | 0.0004 | 0.0015 | 0.0713 |
| CZE | 0.0662 | 0.2650 | 0.4598 | 0.0170 | 0.0110 | 0.8200 |
| DEU | 0.3922 | 0.5318 | 3.7677 | 0.0643 | 0.0403 | 4.7962 |
| DNK | 0.0584 | 0.0281 | 0.2076 | 0.0024 | 0.0066 | 0.3021 |
| ESP | 0.5200 | 0.3517 | 2.0243 | 0.0528 | 0.0328 | 2.9816 |
| EST | 0.0125 | 0.0261 | 0.0597 | 0.0005 | 0.0025 | 0.1013 |
| FIN | 0.0698 | 0.1559 | 0.2776 | 0.0041 | 0.0092 | 0.5166 |
| FRA | 0.0729 | 0.3063 | 3.0583 | 0.0331 | 0.0454 | 3.5160 |
| GBR | 0.0600 | 0.3649 | 2.5213 | 0.0179 | 0.0775 | 3.0416 |
| GRC | 0.1727 | 0.2083 | 0.4222 | 0.0237 | 0.0134 | 0.8402 |
| HRV | 0.0206 | 0.0321 | 0.1333 | 0.0011 | 0.0029 | 0.1890 |
| HUN | 0.0667 | 0.0490 | 0.2913 | 0.0026 | 0.0096 | 0.4193 |
| IRL | 0.0477 | 0.0235 | 0.1746 | 0.0006 | 0.0063 | 0.2527 |
| ISL | 0.0042 | 0.0125 | 0.0214 | 0.0002 | 0.0016 | 0.0398 |
| ITA | 0.5327 | 0.5735 | 3.1215 | 0.1017 | 0.0500 | 4.3794 |
| LTU | 0.0097 | 0.0254 | 0.1144 | 0.0005 | 0.0040 | 0.1539 |
| LUX | 0.0052 | 0.0053 | 0.0332 | 0.0004 | 0.0010 | 0.0452 |
| LVA | 0.0086 | 0.0173 | 0.0566 | 0.0004 | 0.0024 | 0.0853 |
| MNE | 0.0020 | 0.0027 | 0.0165 | 0.0001 | 0.0007 | 0.0221 |
| MKD | 0.0047 | 0.0057 | 0.0332 | 0.0002 | 0.0016 | 0.0454 |
| NLD | 0.1355 | 0.0617 | 0.6756 | 0.0097 | 0.0048 | 0.8862 |
| NOR | 0.0737 | 0.0644 | 0.2204 | 0.0029 | 0.0078 | 0.3692 |
| POL | 0.3974 | 0.6890 | 1.5929 | 0.0225 | 0.0450 | 2.7468 |
| PRT | 0.1901 | 0.0519 | 0.4226 | 0.0067 | 0.0077 | 0.6790 |
| ROU | 0.1130 | 0.1424 | 0.5160 | 0.0019 | 0.0259 | 0.7993 |
| SRB | 0.0291 | 0.0351 | 0.1600 | 0.0006 | 0.0049 | 0.2297 |
| SWE | 0.0855 | 0.0706 | 0.3896 | 0.0045 | 0.0072 | 0.5574 |
| SVN | 0.0140 | 0.0169 | 0.0915 | 0.0011 | 0.0014 | 0.1249 |
| SVK | 0.0373 | 0.0450 | 0.1857 | 0.0017 | 0.0047 | 0.2744 |
| BIH | 0.0087 | 0.0213 | 0.0735 | 0.0002 | 0.0026 | 0.1062 |

**Supplementary Table 5:** Total distance traveled per vehicle segment and country. Unit of measurement: vehicle-kilometer [Mvkm]

| Country | Light-duty | Heavy-duty | Passenger cars | Motorcycles | Buses |
|---|---|---|---|---|---|
| ALB | 1501 | 496 | 6015 | 112 | 232 |
| AUT | 6677 | 4057 | 72129 | 1493 | 572 |
| BEL | 12736 | 3842 | 97227 | 1882 | 658 |
| BGR | 3387 | 1429 | 32491 | 621 | 530 |
| BIH | 2116 | 693 | 9459 | 197 | 331 |
| CHE | 5707 | 2099 | 60099 | 4486 | 350 |
| CYP | 1679 | 87 | 5339 | 123 | 211 |
| CZE | 8664 | 5111 | 50061 | 1597 | 842 |
| DEU | 32930 | 41789 | 549744 | 11118 | 3615 |
| DNK | 7775 | 2257 | 42058 | 683 | 828 |
| ESP | 33921 | 17898 | 291331 | 16358 | 1958 |
| EST | 970 | 310 | 8292 | 63 | 223 |
| FIN | 7745 | 2124 | 42650 | 1051 | 726 |
| FRA | 123951 | 21220 | 468277 | 18159 | 3845 |
| GBR | 79092 | 20111 | 386423 | 4371 | 2347 |
| GRC | 10799 | 2468 | 37966 | 6904 | 946 |
| HRV | 1910 | 1018 | 22730 | 202 | 382 |
| HUN | 7412 | 2758 | 34978 | 478 | 898 |
| IRL | 5835 | 2145 | 36461 | 96 | 492 |
| ITA | 59508 | 9733 | 381739 | 30650 | 4636 |
| LTU | 1095 | 915 | 20935 | 101 | 108 |
| LUX | 607 | 217 | 6164 | 64 | 63 |
| LVA | 1136 | 626 | 9374 | 70 | 164 |
| MKD | 1136 | 371 | 5505 | 109 | 177 |
| MLT | 559 | 156 | 3009 | 45 | 106 |
| MNE | 432 | 142 | 1860 | 37 | 67 |
| NLD | 20411 | 6289 | 107549 | 4639 | 711 |
| NOR | 5639 | 2079 | 30214 | 259 | 653 |
| POL | 43219 | 25551 | 191785 | 3077 | 3124 |
| PRT | 12947 | 2015 | 64903 | 1480 | 409 |
| ROU | 11516 | 1626 | 57706 | 560 | 2855 |
| SRB | 3518 | 1127 | 17151 | 468 | 547 |
| SVK | 3593 | 2723 | 16719 | 101 | 412 |
| SVN | 1576 | 881 | 18501 | 227 | 154 |
| SWE | 6740 | 3229 | 57890 | 560 | 719 |

**Supplementary Table 6:** Number of vehicles per country (vehicle number). Source: Mantzos et al. (2018)

| Country | Light-duty | Heavy-duty | Passenger cars | Motorcycles | Buses |
|---|---|---|---|---|---|
| AUT | 422,745 | 53,582 | 4,978,852 | 536,964 | 10,037 |
| ALB | 60,036 | 10,864 | 460,027 | 24,498 | 7,146 |
| BEL | 759,406 | 96,690 | 5,853,782 | 490,495 | 16,125 |
| BGR | 348,667 | 38,519 | 2,773,325 | 112,388 | 20,818 |
| CHE | 374,819 | 42,174 | 4,666,278 | 739,344 | 14,395 |
| CYP | 98,533 | 10,511 | 550,695 | 28,183 | 3,084 |
| CZE | 441,303 | 264,959 | 5,747,913 | 1,132,085 | 22,027 |
| DEU | 2,616,118 | 531,849 | 47,095,784 | 4,284,918 | 80,519 |
| DNK | 389,461 | 28,124 | 2,594,482 | 160,823 | 13,158 |
| ESP | 3,466,479 | 351,702 | 25,304,190 | 3,520,005 | 65,599 |
| EST | 83,313 | 26,102 | 746,464 | 34,629 | 5,026 |
| FIN | 465,024 | 155,868 | 3,470,507 | 272,168 | 18,467 |
| FRA | 4,854,678 | 306,298 | 38,229,305 | 2,205,578 | 90,807 |
| GBR | 4,003,781 | 364,877 | 31,517,597 | 1,193,841 | 154,990 |
| GRC | 1,151,020 | 208,280 | 5,282,695 | 1,583,491 | 26,743 |
| HRV | 137,049 | 32,126 | 1,666,413 | 73,997 | 5,877 |
| HUN | 444,588 | 49,017 | 3,641,823 | 176,070 | 19,134 |
| IRL | 317,798 | 23,526 | 2,182,920 | 38,260 | 12,500 |
| ISL | 28,065 | 12,484 | 267,467 | 10,832 | 3,146 |
| ITA | 3,556,816 | 573,475 | 39,018,170 | 6,780,733 | 100,042 |
| LTU | 64,345 | 25,409 | 1,430,520 | 33,666 | 7,925 |
| LUX | 34,833 | 5,348 | 415,145 | 24,054 | 2,042 |
| LVA | 57,146 | 17,283 | 707,841 | 23,713 | 4,885 |
| MNE | 13,440 | 2,718 | 206,453 | 5,655 | 1,459 |
| MKD | 31,339 | 5,671 | 415,062 | 13,343 | 3,201 |
| NLD | 903,005 | 61,652 | 8,442,982 | 646,046 | 9,513 |
| NOR | 491,649 | 64,365 | 2,751,949 | 192,536 | 15,634 |
| POL | 2,649,198 | 688,968 | 19,911,107 | 1,502,888 | 89,905 |
| PRT | 1,267,647 | 51,908 | 5,282,970 | 447,876 | 15,493 |
| ROU | 753,029 | 142,414 | 6,452,536 | 129,657 | 51,802 |
| SRB | 193,827 | 35,073 | 1,999,771 | 40,695 | 9,880 |
| SWE | 570,252 | 70,617 | 4,869,979 | 300,356 | 14,377 |
| SVN | 93,484 | 16,916 | 1,143,150 | 67,145 | 2,834 |
| SVK | 248,873 | 45,034 | 2,321,608 | 110,141 | 9,363 |
| BIH | 57,805 | 21,268 | 919,226 | 10,552 | 5,212 |

**Supplementary Table 7:** Estimated battery capacity per country and transport segment in 2050 (TWh). We assume battery capacities of 80 kWh for passenger vehicles, 150 kWh for light-duty vehicles, 15 kWh for motorcycles, 500 kWh for buses, and 1000 kWh for heavy-duty vehicles.

| Country | Light-duty | Heavy-duty | Passenger car | Motorcycle | Bus | Total |
|---|---|---|---|---|---|---|
| AUT | 0.0634 | 0.0536 | 0.3983 | 0.0080 | 0.0050 | 0.5284 |
| ALB | 0.0090 | 0.0109 | 0.0368 | 0.0004 | 0.0036 | 0.0606 |
| BEL | 0.1139 | 0.0967 | 0.4683 | 0.0074 | 0.0081 | 0.6943 |
| BGR | 0.0523 | 0.0385 | 0.2219 | 0.0017 | 0.0104 | 0.3248 |
| CHE | 0.0562 | 0.0422 | 0.3733 | 0.0111 | 0.0072 | 0.4900 |
| CYP | 0.0148 | 0.0105 | 0.0441 | 0.0004 | 0.0015 | 0.0713 |
| CZE | 0.0662 | 0.2650 | 0.4598 | 0.0170 | 0.0110 | 0.8200 |
| DEU | 0.3922 | 0.5318 | 3.7677 | 0.0643 | 0.0403 | 4.7962 |
| DNK | 0.0584 | 0.0281 | 0.2076 | 0.0024 | 0.0066 | 0.3021 |
| ESP | 0.5200 | 0.3517 | 2.0243 | 0.0528 | 0.0328 | 2.9816 |
| EST | 0.0125 | 0.0261 | 0.0597 | 0.0005 | 0.0025 | 0.1013 |
| FIN | 0.0698 | 0.1559 | 0.2776 | 0.0041 | 0.0092 | 0.5166 |
| FRA | 0.0729 | 0.3063 | 3.0583 | 0.0331 | 0.0454 | 3.5160 |
| GBR | 0.0600 | 0.3649 | 2.5213 | 0.0179 | 0.0775 | 3.0416 |
| GRC | 0.1727 | 0.2083 | 0.4222 | 0.0237 | 0.0134 | 0.8402 |
| HRV | 0.0206 | 0.0321 | 0.1333 | 0.0011 | 0.0029 | 0.1890 |
| HUN | 0.0667 | 0.0490 | 0.2913 | 0.0026 | 0.0096 | 0.4193 |
| IRL | 0.0477 | 0.0235 | 0.1746 | 0.0006 | 0.0063 | 0.2527 |
| ISL | 0.0042 | 0.0125 | 0.0214 | 0.0002 | 0.0016 | 0.0398 |
| ITA | 0.5327 | 0.5735 | 3.1215 | 0.1017 | 0.0500 | 4.3794 |
| LTU | 0.0097 | 0.0254 | 0.1144 | 0.0005 | 0.0040 | 0.1539 |
| LUX | 0.0052 | 0.0053 | 0.0332 | 0.0004 | 0.0010 | 0.0452 |
| LVA | 0.0086 | 0.0173 | 0.0566 | 0.0004 | 0.0024 | 0.0853 |
| MNE | 0.0020 | 0.0027 | 0.0165 | 0.0001 | 0.0007 | 0.0221 |
| MKD | 0.0047 | 0.0057 | 0.0332 | 0.0002 | 0.0016 | 0.0454 |
| NLD | 0.1355 | 0.0617 | 0.6756 | 0.0097 | 0.0048 | 0.8862 |
| NOR | 0.0737 | 0.0644 | 0.2204 | 0.0029 | 0.0078 | 0.3692 |
| POL | 0.3974 | 0.6890 | 1.5929 | 0.0225 | 0.0450 | 2.7468 |
| PRT | 0.1901 | 0.0519 | 0.4226 | 0.0067 | 0.0077 | 0.6790 |
| ROU | 0.1130 | 0.1424 | 0.5160 | 0.0019 | 0.0259 | 0.7993 |
| SRB | 0.0291 | 0.0351 | 0.1600 | 0.0006 | 0.0049 | 0.2297 |
| SWE | 0.0855 | 0.0706 | 0.3896 | 0.0045 | 0.0072 | 0.5574 |
| SVN | 0.0140 | 0.0169 | 0.0915 | 0.0011 | 0.0014 | 0.1249 |
| SVK | 0.0373 | 0.0450 | 0.1857 | 0.0017 | 0.0047 | 0.2744 |
| BIH | 0.0087 | 0.0213 | 0.0735 | 0.0002 | 0.0026 | 0.1062 |

# References

Burke, A. F., Zhao, J., and Fulton, L. M. (2024). Projections of the costs of light-duty battery-electric and fuel cell vehicles (2020–2040) and related economic issues. *Research in Transportation Economics*, 105:101440.

Danish Energy Agency (2024). Technology catalogue for commercial freight and passenger transport. Accessed: 2025-09-24.

Herbst, I., Barth, J., Daneshgar, G., Furrer, R., Heer, L., Hickethier, M., Mathis, U., Rosser, S., Trottmann, M., Vischer, M., Wahl, H., and Walter, S. (2023). Laden im quartier: Informationssammlung zur elektromobilität für gemeinden. https://pubdb.bfe.admin.ch/de/publication/download/12068. PDF verfügbar unter https://pubdb.bfe.admin.ch/de/publication/download/12068.

Lombardi, F., van Greevenbroek, K., Grochowicz, A., Lau, M., Neumann, F., Patankar, N., and Vågerö, O. (2025). Near-optimal energy planning strategies with modeling to generate alternatives to flexibly explore practically desirable options. *Joule*, 9(11). Publisher: Elsevier.

Mantzos, L., Rozsai, M., Matei, N. A., Mulholland, E., Tamba, M., and Wiesenthal, T. (2018). Jrc-idees 2015.

Pasaoglu, G., Fiorello, D., Martino, A., Scarcella, G., Alemanno, A., Zubaryeva, A., and Thiel, C. (2012). Driving and parking patterns of european car drivers: A mobility survey. JRC Scientific and Policy Report JRC77079, EUR 25627 EN, Joint Research Centre, European Commission, Petten, Netherlands / Luxembourg.

Schwarzer, C. M. (2025). Bidirectional charging: How an electric car (finally) earns money. Accessed: 2026-01-20.

Wood, E., Borlaug, B., Moniot, M., Lee, D.-Y., Ge, Y., Yang, F., and Liu, Z. (2023). The 2030 national charging network: Estimating u.s. light-duty demand for electric vehicle charging infrastructure. Technical Report NREL/TP-5400-85654, National Renewable Energy Laboratory.